\documentclass[11pt]{JHEP3}
\usepackage{epsfig}
\usepackage{amsmath,amssymb, amsfonts}
\usepackage{graphicx}
\usepackage{comment}

\def\braket#1{\mathinner{\langle{#1}\rangle}}
\newcommand{\sbraket}[1]{\lbrack #1\rbrack}

\renewcommand{\a}{\alpha}
\renewcommand{\b}{\beta}
\newcommand{\da}{\dot{\alpha}}
\newcommand{\db}{\dot{\beta}}

\newtheorem{suspicion}{Suspicion}[section]

\newcommand{\boxit}[1]{%
  \[\fbox{%
      \addtolength{\linewidth}{-2\fboxsep}%
      \addtolength{\linewidth}{-2\fboxrule}%
      \begin{minipage}{\linewidth}%
      #1%
      \end{minipage}%
    } \nonumber \]%
}

\title{On BCFW shifts of integrands and integrals}
\author{Rutger Boels \\  II. Institut f\"ur Theoretische Physik Universit\"at Hamburg\\ Luruper Chaussee 149, D- 22761 Hamburg, Germany }

\abstract{In this article a first step is made towards the extension of Britto-Cachazo-Feng-Witten (BCFW) tree level on-shell recursion relations to integrands and integrals of scattering amplitudes to arbitrary loop order. Surprisingly, it is shown that the large BCFW shift limit of the integrands has the same structure as the corresponding tree level amplitude in any minimally coupled Yang-Mills theory in four or more dimensions. This implies that these integrands can be reconstructed from a subset of their `single cuts'. The main tool is powercounting Feynman graphs in a special lightcone gauge choice employed earlier at tree level by Arkani-Hamed and Kaplan. The relation between shifts of integrands and shifts of its integrals is investigated explicitly at one loop. Two particular sources of discrepancy between the integral and integrand are identified related to UV and IR divergences. This is cross-checked with known results for helicity equal amplitudes at one loop. The nature of the on-shell residue at each of the single-cut singularities of the integrand is commented upon. Several natural conjectures and opportunities for further research present themselves.}

\begin{document}

\section{Introduction}
Scattering amplitudes are on the crossroads of many developments in physics and are therefore prime objectives for the developments of new calculational methods. In recent years the technology for analytic calculation of these amplitudes has seen many advances inspired by Witten's twistor insights \cite{witten}. Contact of these developments with the calculation of experimentally (still!) relevant NLO calculations is being made. However, many of the recently developed techniques are applicable for QCD only at the tree or one-loop level, while many of the most exciting analytic developments have been at high loop level in $\mathcal{N}=4$ super Yang-Mills theory. NNLO calculations in QCD are certainly on the (longer term) wishlist of experimenters. Apart from this direct motivation from the experimental side there is the inherent intuition that many of the recently found hidden structures in maximally supersymmetric Yang-Mills might have some form of non-supersymmetric counterparts. In this article the full extension of one particular tree level technique to the loop level at in principle arbitrary loop order will be initiated. This technique will be that of the Britto-Cachazo-Feng-Witten (BCFW) on-shell recursion relations \cite{Britto:2004ap}, \cite{Britto:2005fq}. In a nutshell, at tree level these relations allow one to calculate any amplitude from the analytic form of the three point amplitude only. There is a broad parallel here to bootstrap equations in CFT as pointed out in \cite{Boels:2010bv}. This looks especially natural from a twistor transformed point of view \cite{Mason:2009sa}, \cite{ArkaniHamed:2009si}.  From the CFT bootstrap perspective it is natural to widen the investigation of recursion relations to the loop level. 

On-shell recursion at loop level for complete amplitudes has certainly been discussed previously in several places in the literature, starting in \cite{Bern:2005hs}, where the purely rational $1$-loop amplitudes in pure Yang-Mills were studied starting from their known expressions. In further work this has been developed into a method for calculating rational terms \cite{Bern:2005cq}, \cite{Berger:2006ci} in general which has in addition been implemented in numerical code \cite{Berger:2008sj}. Recursion for coefficients of the loop integrals has also been studied \cite{Bern:2005hh} and a recent application of these techniques includes for instance the calculation of rational terms in some pure Einstein one loop gravity amplitudes \cite{Dunbar:2010xk}. One technical problem in all these approaches is in general the appearance of so-called boundary contributions for large BCFW shifts which spoil direct predictability. Controlling these can be done using auxiliary recursion relations  \cite{Bern:2005cq}, \cite{Berger:2006ci} and general powerful consistency conditions but a direct understanding of the boundary contributions has been missing so far. Boundary contributions appear also in other context, see e.g. \cite{Feng:2009ei}, \cite{Feng:2010ku} for a very recent discussion. Moreover all of the mentioned work depends on properties of amplitudes special to one loop.

In this article the known issues with on-shell recursion for amplitudes at loop level are first sidestepped by considering the \emph{integrand} of the amplitudes instead of the integrated expressions. It is quite straightforward to check in several explicitly known examples of amplitudes in gauge theories that the scaling of the integrand matches that of the corresponding tree level amplitudes. This can be seen in for instance the case of MHV amplitudes in $\mathcal{N}=4$ at one loop as obtained by long ago \cite{Bern:1994zx} and the all-plus amplitudes in pure Yang-Mills, also at one loop (using the fact that these amplitudes are related \cite{Bern:1996ja}). By inspection the full integrand of these amplitudes is seen to behave under BCFW shifts exactly as the tree amplitude. The same result also follows for the integrals appearing in the planar $4$ point amplitude in $\mathcal{N}=4$ at five loops as given in \cite{Bern:2007ct}. This leads to the following suspicion:

\begin{suspicion}\label{sus:main}
In four dimensions and up, as long as all loop momenta and the external momenta are sufficiently generic, the integrand of Yang-Mills and Einstein gravity amplitudes coupled to fundamental or adjoint matter scales under a BCFW shift of a pair of outside legs as the corresponding tree amplitudes to all loop orders. 
\end{suspicion}

Although the focus in this article will be on gauge theory, Einstein gravity is included in the above suspicion since this is quite natural to suspect. For gauge theory there is one class of shifts which does not exist at tree level: those of gluons on different color traces. One expects these to behave as non-adjacent shifts at tree level, i.e. they should be more (at least one power of $1/z$) suppressed under large shifts. As explained below the shift behavior proves the integrand can be reconstructed from the propagator pole singularities in the shift parameter. If the residues at the poles of the integrand can be interpreted in terms of amplitudes again this implies 

\begin{suspicion}\label{sus:main2}
In four dimensions and up, as long as all loop momenta and the external momenta are sufficiently generic, the integrand of Yang-Mills and Einstein gravity amplitudes coupled to fundamental or adjoint matter obeys (a form of) on-shell recursion relations to all loop orders. 
\end{suspicion}

The residues at the poles in the integrand correspond to single cuts: replacing one particular propagator by a delta function. These single cuts of the integrand are reminiscent of but distinct from Feynman's tree theorem \cite{Feynman:1972mt} or even more the streamlined versions of the tree theorem in \cite{Catani:2008xa},  \cite{Bierenbaum:2010cy} and especially \cite{CaronHuot:2010zt}. The difference is that the construction studied in this paper will turn out to involve a certain strict subset of single cuts only while keeping propagators the same as usual. In contrast, the mentioned applications of Feynman's tree theorem study all cuts in the original form or all single cuts in the more advanced version with certain modified propagators (or boundary conditions). Note that for direct numerical applications the approach of Catani et. al. might be better suited as it treats all singularities equally, allowing in principle to treat all the IR singularities of real and virtual corrections on an even footing. The usefulness of applying a more conventional expand-in-a-basis unitarity-type approach using single particle cuts has been studied in \cite{NigelGlover:2008ur}.

From the outset it is clear that a tension exists between the limit of large BCFW shift of the integrands versus that of the integrals. This arises since the integrals are only well-defined after (dimensional) regularization, so this can be viewed as an order of limits problem. For the leading poles in the dimensional regularization parameter $\epsilon$ this is not expected to be much of a problem, but for sub-leading terms a problem might arise. At one loop this is a well-known phenomenon related to the appearance of so-called 'rational terms'. This is motivation to study the difference between integrands and integrals more closely as will be done below. 

This article is structured as follows: in section \ref{sec:bcfwtrees} the standard tree level BCFW recursion relations are first reviewed. For the derivation the large BCFW shift behavior of the tree level amplitude is crucial and a completely diagrammatic version of this derivation is presented using powercounting in a special lightcone gauge. This argument is independent of much of the fine details of the exact correlation function under study such as the loop order. This is expanded on further in section \ref{sec:bcfwloopsintegr}. In particular, the diagrammatic argument shows that the integrand of Yang-Mills theory coupled to various kinds of matter has the same BCFW shift as the tree level amplitude for shifts of particles on the same color trace thereby establishing suspicion \ref{sus:main} for gauge theory for this class of shifts. This result on integrands begs the question how the shift relates to the integrals of the integrands. This question is studied first at one loop in section \ref{sec:integralsvsintegrands} using the background field method to separate gauge choices of trees and loops. In particular the large BCFW shifts of the purely rational amplitudes at one loop are studied. In section \ref{sec:reltoamp} a preliminary discussion of the relation of the single cuts of the integrand to lower point or lower loop amplitudes is given as a first step in the direction towards suspicion \ref{sus:main2}. A general discussion and conclusion section rounds off the presentation. In appendix \ref{app:BCFWconv} shifts of other particles than gluons are considered in four and higher dimensions. Appendix \ref{app:loopintegrals} contains a set of explicit results for a series of Feynman graphs used in section \ref{sec:reltoamp}. 

\textit{Note added in proof:} Simultaneously with this article closely related and independent work by Arkani-Hamed et.al.  \cite{ArkaniHamed:2010kv} has appeared on the archive which contains an explicit example of a recursion relation for the integrand for planar $\mathcal{N}=4$ super Yang-Mills theory in four dimensions.

\section{BCFW recursion for tree level amplitudes}
\label{sec:bcfwtrees}
As BCFW observed amplitudes can be turned into a function of a single complex variable by shifting the momenta of two legs of the amplitude as
\begin{equation}\label{eq:BCFWshift}
\begin{array}{cc}k_i \rightarrow & k_i + q z \\
k_{j} \rightarrow  & k_j - q z \ ,
\end{array}
\end{equation}
while requiring the vector $q$ to obey
\begin{equation}\label{eq:BCFWconstraint} 
q \cdot k_i = q\cdot k_j = q \cdot q = 0 \, \ .
\end{equation}
which keeps the invariant lengths  $k_i^2$ and $k_j^2$ invariant. These equations have two complex valued solutions for $q$, as can be checked in an appropriate (`center of mass') Lorentz-frame. A covariant solution can also be constructed using higher dimensional spinor helicity methods \cite{Cheung:2009dc} \cite{Boels:2009bv}, see appendix \ref{app:BCFWconv} for details. The shift turns the amplitude into a function of the complex shift parameter $z$. This function can then be contour integrated to yield the amplitude of interest,
\begin{equation}
A(0) = \oint_{z=0} \frac{A(z)}{z}  \ .
\end{equation}

Up to this point no assumption about the loop order of the amplitude has been made up to the reasonable assumption of the singularity at $z=0$ to be a single simple pole. If the amplitude is tree level however, it must be a rational function of the momenta and hence also of $z$. In this case one can pull the contour to infinity as depicted in figure \ref{fig:BCFWtrees} to yield\footnote{Here and in the following the $\oint$ symbol stands for the integral $\frac{1}{2 \pi i} \oint dz$.}
\begin{equation}\label{eq:BCFWrecursionint}
A(0) = \oint_{z=0} \frac{A(z)}{z} = \sum_{res} \left(z = \textrm{finite}\right) +  \sum_{res} \left(z = \textrm{infinite}\right) \ .
\end{equation}
 In this formula the finite $z$ poles have known residues by tree level unitarity. These residues are a product of tree amplitudes, summed over all states in the cut channel and summed over all factorization channels where the shifted particles are on distinct tree amplitudes in the residue (possibly taking into account color ordering). Importantly, the involved amplitudes have a strictly smaller number of particles. The infinite $z$ residues do not have a similar interpretation. If the latter residues are absent however, equation \eqref{eq:BCFWrecursionint} is an on-shell recursion relation.

\FIGURE[h]{\epsfig{file=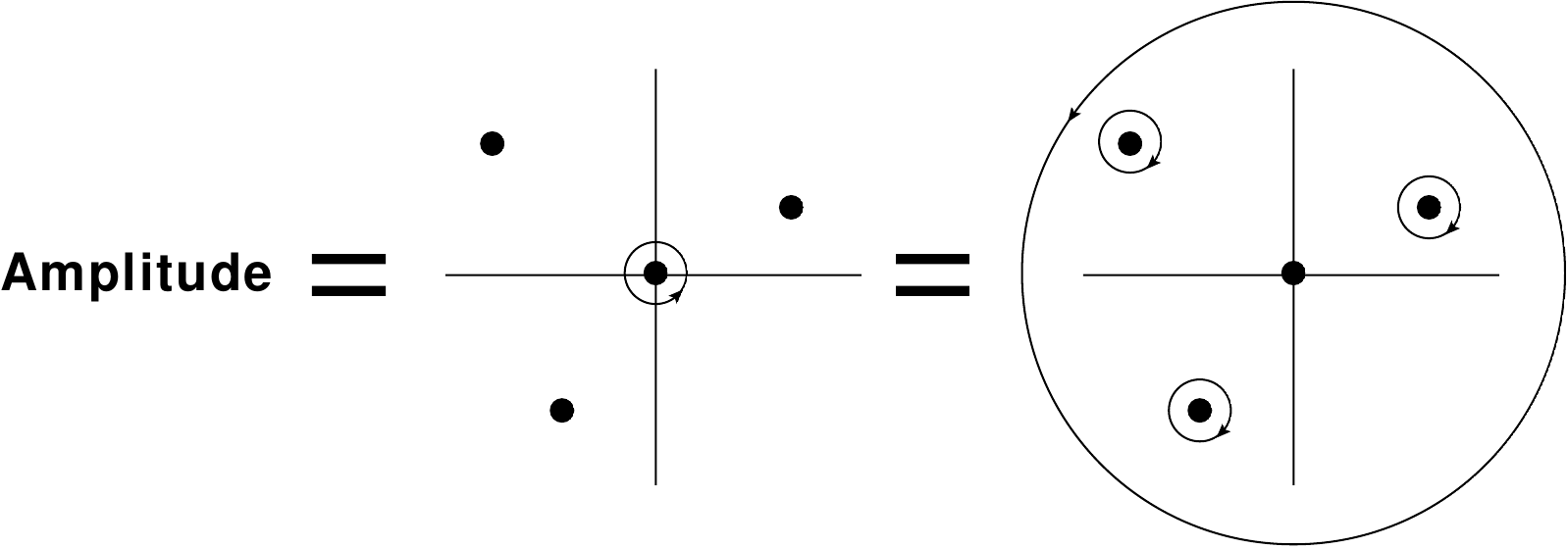,scale=0.7} \caption{Sketch of contours and singularities used for BCFW recursion at tree level in equation \ref{eq:BCFWrecursionint}} \label{fig:BCFWtrees}}

The challenge is therefore to derive the large $z$-behavior to show the absence of these residues. In general this depends on the helicity of the particles being shifted and whether or not the shifted particles are adjacent in the color ordering (for gauge theories). For  gluons in Yang-Mills theory the large $z$-behavior is for instance displayed in table \ref{tab:largezn4} for shifts of particles adjacent in the color ordering. In this table the difference between $T$ and $T2$ is whether the inner product between the two polarization vector vanishes or not. If it does, the shift is better behaved. Non-color-adjacent shifts are suppressed by an additional factor of $\left(\frac{1}{z}\right)$. See table \ref{tab:largezferms} in particular and appendix \ref{app:BCFWconv} in general for how amplitudes behave for shifts of gluonic and fermionic legs. .

\begin{table}
\begin{center}
\begin{tabular}{c|c c c}
$\epsilon_1 \;\backslash \;\epsilon_2  $ & $-$              & $+$              & T \\
\hline
$-$                   & $ +1$ & $ +1$ & $ +1$ \\
$+$                   & $ -3$ & $ +1$ & $ -1$ \\
T                     & $ -1$ & $ +1$ & $ -1$ \\
T2                    & $ -1$ & $ +1$ & $0$ \\
\end{tabular}

\caption{leading asymptotic power in $z^{-\kappa}$ of the adjacent BCFW shift of two gluons in a  tree amplitude for all possible polarizations of the shifted gluons}
 \label{tab:largezn4}
\end{center}
\end{table}

Table \ref{tab:largezn4} can be derived through various means but perhaps most physical Arkani-Hamed and Kaplan \cite{ArkaniHamed:2008yf} pointed out that a large BCFW shift basically describes a hard particle shooting through a soft background given by the other gluons (see equation \ref{eq:BCFWshift}). Technically, this can be described handily by the background field method. The background field can be put in the $q$ lightcone gauge to show that the naive powercounting of the large $z$-behavior of diagrams in which the shifted legs are on different vertices is suppressed by $\left(\frac{1}{z^2}\right)$. This leaves local vertices which can be analyzed by expanding the Yang-Mills action coupled to for instance fermions to second order in the `hard' fields using this method gives 
\begin{equation}
\mathcal{L} = D_{\nu} a_{\mu}  D^{\nu} a^{\mu} + a_{\mu} a_{\nu} F^{\mu\nu}[A]  + a_{\mu} \left( \overline{\Psi}\gamma^{\mu} \psi + \overline{\psi}\gamma^{\mu} \Psi \right) + A_{\mu} \overline{\psi}\gamma^{\mu} \psi \ , 
\end{equation}
where $a_{\mu}$ and $\psi$ describe the hard particles, while $A^{\mu}$ and $\Psi$ describe the background field. $D^{\nu}$ is the covariant background field derivative. The linear expansion term,
\begin{equation}
\sim a_{\mu} \left(\frac{\delta}{\delta {A_{\mu}}} \mathcal{L}[A,\Psi]  \right) + \psi \left(\frac{\delta}{\delta {\Psi}}  \mathcal{L}[A,\Psi]  \right) \ , 
\end{equation}
vanishes by the equation of motion for the soft fields. The structure of this Lagrangian immediately leads to a prediction of the large $z$ structure of the two particle current which in turn yields table \ref{tab:largezn4} when paired with the external wave functions. 

\subsection{Powercounting Feynman graphs in a special gauge}
The original AHK argument depends on an application of the tree level equation of motion for the external fields. Below we show one can reproduce the same argument in a more diagrammatic language, which will be useful below. The BCFW on-shell recursion relations have been studied from the diagrammatic point of view before in \cite{Draggiotis:2005wq}, but there the focus was on the much harder task of understanding the terms in the recursion relations directly. Here the focus will be firmly on the large BCFW limit. As a bonus explicit expressions for the first and second non-vanishing coefficients in the large $z$ limit will be produced. For this the amplitude under study is computed in what will be referred to as the Arkani-Hamed-Kaplan (AHK) gauge,
\begin{equation}
q_{\mu} A^{\mu} = 0 \, \ .
\end{equation}
The corresponding lightcone gauge propagator reads
\begin{equation}\label{eq:lcgprop}
G(k)_{\mu\nu} \sim \frac{1}{k^2} \left[\eta_{\mu\nu} - \left(\frac{q_{\mu} k_{\nu} + k_{\mu} q_{\nu} }{k \cdot q} \right)\right] \, \ ,
\end{equation}
with $\eta$ the flat space metric. This propagator connects the usual (color ordered if required) three and four point vertices and the external fields. The auxilliary field is left in the Lagrangian in contradistinction to the usual lightcone approach. For our purposes the external wave functions of the shifted legs will be kept in a different gauge than the other legs which will be put in $q$ lightcone gauge. This is necessary since in the large $z$ limit the momenta of the shifted legs become proportional to $q$ and the AHK gauge is therefore singular for these two special external wave functions. 

There is a more serious gauge singularity for those diagrams for which the shifted legs end on the same three vertex. For this special class $q$ is orthogonal to the momentum in the soft leg,
\begin{equation}
q \cdot (k_1 + k_2) = 0 \ , 
\end{equation}
and the above lightcone gauge propagator in equation \eqref{eq:lcgprop} is singular. This difficulty can be circumvented by first imposing the auxilliary lightcone gauge
\begin{equation}
(q+ x k_1)_{\mu} A^{\mu} = 0 \ .
\end{equation}
For the diagram which is singular in the limit $x \rightarrow 0$ the relevant propagator shows of course an explicit pole in $x$,
\begin{equation}
G(k_1 + k_2)_{\mu\nu} = \frac{1}{(k_1 + k_2)^2} \left(\eta_{\mu\nu} - \frac{(q+ x k_1)_{\mu} (k_1+k_2)_{\nu} + (k_1+k_2)_{\mu}(q+ x k_1)_{\nu} }{x (k_1 \cdot k_2)} \right) \ .
\end{equation}
This pole should be a gauge artefact since the complete amplitude is gauge invariant. This can be made more precise. From the usual expression of the (color ordered) three vertex in Yang-Mills the contribution of the diagrams with the shifted legs ending on the same three vertex coupled to an arbitrary remainder through the above lightcone propagator can be written as 
\begin{align}\label{eq:largezsingcontrib}
\epsilon_{1}^{\mu}\left(\hat{k}_1\right) & \epsilon_{2}^{\nu}\left(\hat{k}_2\right) V_{\mu\nu}^{\kappa}  \sim \nonumber \\
& \sim   \epsilon_{1}^{\mu} \epsilon_{2}^{\nu}  \left[\eta_{\nu\mu} (2 z q + k_1 - k_2)^{\rho} + (\eta_{\mu}^{\rho} (k_1+k_2)_{\nu} - \eta_{\nu}^{\rho} (k_1+k_2)_{\mu}) \right] G_{\rho}^{\kappa} \nonumber\\ 
  & \sim  \frac{\epsilon_{1}^{\mu} \epsilon_{2}^{\nu}}{(k_1 + k_2)^2} \bigg[\eta_{\nu\mu} (2 z q - 2 k_1)^{\kappa} + (\eta_{\mu}^{\kappa} (k_1+k_2)_{\nu} - \eta_{\nu}^{\kappa} (k_1+k_2)_{\mu}) +\nonumber  \\ & \qquad \qquad \left. \frac{\left[(q)_{\mu} (k_1 + k_2)_{\nu} - (q + x k_1)_{\nu} (k_1 + k_2)_{\mu} \right] (k_1 + k_2)^{\kappa})}{x (k_1 \cdot k_2 )}\right]   \ ,
\end{align}
where the hats on the momenta indicate BCFW shifted quantities. To obtain this one uses
\begin{align}\label{eq:specprops}
q^{\rho} & G(k_1 + k_2)_{\rho}^{\kappa} = \frac{q^{\kappa} }{(k_1 + k_2)^2}  \\
(k_1 - k_2)^{\rho} & G(k_1 + k_2)_{\rho}^{\kappa} = -\frac{2 k_{1}^{\kappa}}{(k_1 + k_2)^2} \ .
\end{align}
All other (parts of the) diagrams in the theory do not have poles in $x$ for sufficiently generic momenta. Hence one expects the poles on the last line of equation \eqref{eq:largezsingcontrib} to vanish. Actually, since the possible pole arises from either
\begin{equation}
q_{\mu} e_1^{\mu} \quad \textrm{or} \quad q_{\mu} e_2^{\mu} \ , 
\end{equation}
it is seen that the possible pole in $x$ is lower order in $z$ since
\begin{equation}\label{eq:simplcontractwithq}
q_{\mu} e_i^{\mu} = \pm_i \frac{k_{i,\mu}}{z} e_i^{\mu}  \qquad i=1,2 \ ,
\end{equation}
where the sign depends on the particle. Hence for our purposes the terms involving $\frac{1}{x}$ poles can be consistently dropped. Moreover, the entire pole at $x=0$ vanishes by gauge invariance of the one particle off-shell current which contracts into the above vertex. This follows as the momentum of the off-shell leg is $(k_1 + k_2)$.

Hence the class of diagrams where the shifted legs end on the same vertex contributes terms of the form
\begin{equation}
A(z) \rightarrow \epsilon_{1}^{\mu}\left(\hat{k}_1\right) \epsilon_{2}^{\nu}\left(\hat{k}_2\right) M_{\mu\nu}  = 
\epsilon_{1}^{\mu}\left(z \right) \epsilon_{2}^{\nu}\left(z\right) \left( z \, \eta_{\mu\nu}   f_1\left(1/z\right) +  f_{2,\mu\nu}\left(1/z\right)+ \mathcal{O}\left(\frac{1}{z}\right) \right) + \ldots \, \ ,
\end{equation}
to the large shift limit. Here $f_{2,\mu\nu}$ is an antisymmetric matrix and the dots indicate the diagrams where the shifted legs end on different vertices to be analyzed below. The structure of the leading and sub-leading terms here is basically the AHK result \cite{ArkaniHamed:2008yf}. The functions $f_i$ are in general polynomials in $\left(\frac{1}{z}\right)$ with non-trivial constant term.  

For the class of diagrams just considered the function $f_1$ for instance can be calculated from the coupling of the one leg off-shell  current to the above three vertex current with two shifted legs as
\begin{equation}\label{eq:f1treelevel}
f_1 = \left(q_{\mu} - \frac{k_{1,\mu}}{z}  \right)J_{\mu}^{q}(X) \ .
\end{equation}
Here $J_{\mu}^{q}$ is the current and $X$ stands for the quantum numbers on the other legs of the diagram. Note that in a real sense the leading-in-$z$ contribution to the large BCFW shift can be thought of as an effective particle made out of particles one and two. This effective particle interpretation follows also from the CFT analysis of BCFW shifts in string theory \cite{Cheung:2010vn} \cite{Boels:2010bv}. In the string context the corresponding particle is sometimes referred to as the pomeron \cite{Brower:2006ea}. 

For the above result the contributions to the amplitude of those diagrams where the shifted legs end on different vertices has been disregarded. This will now be justified. The $z$ dependence of these diagrams can be obtained from a refined form of powercounting. Note that in every tree diagram there is a unique path from one shifted leg to the other. This path will be referred to as the 'hard line' in the following as it is the path along which the $z$ dependence flows. Every three vertex along the hard line scales as $\left(z^1\right)$ with the legs contracted into $q$ and a metric while every four vertex scales as $\left(z^0\right)$ with two metrics contracting the legs. Along the hard line the lightcone gauge propagator scales as 
\begin{multline}\label{eq:AHKgaugeprophardline}
G(k)_{\mu\nu} \sim \frac{1}{k^2} \left[\eta_{\mu\nu} - \left(\frac{q_{\mu} k_{\nu} + k_{\mu} q_{\nu} }{k \cdot q} \right)\right] \rightarrow \\ \frac{1}{k^2 \pm 2 z {q k}} \left[\eta_{\mu\nu} - \left(\frac{q_{\mu} k_{\nu} + k_{\mu} q_{\nu} }{k \cdot q} \right)\right]  \mp \frac{2 z}{k^2 \pm 2 z {q k}} \left[ \left(\frac{q_{\mu} q_{\nu} }{k \cdot q} \right)\right] \ ,
\end{multline}
which contains a term proportional to $z^0$ in the limit. The sign depends on the routing of the momentum. However, although the superficial degree of scaling of a given diagram is now $z^{v_3}$ with $v_3$ the number of three vertices, this is never realized. The above propagator remains after all orthogonal to q, so the part of the three vertex proportional to $q$ never contributes. The order $z$ part of the above propagator has to be contracted with soft field propagators which are orthogonal to q or into the soft external fields also orthogonal to $q$. This leaves the two hard external fields and contractions along the hard line. The only way a possible $z$-dependence could arise from the external hard wave functions would entail terms like
\begin{equation}
\epsilon_{1}^{\mu}(\hat{k}_1) \hat{k}^2_{\mu} =  \epsilon_{1}^{\mu}(\hat{k}_1) \left(\hat{k}^2_{\mu} + \hat{k}^1_{\mu} \right) = \epsilon_{1}^{\mu}(\hat{k}_1) \left(k^2_{\mu} + k^1_{\mu} \right) \ ,
\end{equation}
and its natural conjugate which are therefore suppressed.  Moreover, for this class of contributions there is now an external field contracted into the lightcone gauge propagator along the hard line which leads to a $\left( \frac{1}{z}\right)$ suppression. It is seen that only contracting the momenta on two three vertices using the lightcone propagator and with the shifted fields on the external legs of these three vertices yields an order $z^0$ contribution. In fact, the form of this is easy to derive up to a constant,
\begin{equation}
\epsilon_{1}^{\mu}(\hat{k}_1)  \epsilon_{2}^{\nu}(\hat{k}_2) V^3_{\mu\rho\kappa} G^{\kappa \lambda} V^3_{\lambda \nu \sigma} \sim z^0 \epsilon_{1}^{\mu} \epsilon_{2}^{\nu}  \eta_{\mu\kappa} \eta_{\nu\sigma}+ \mathcal{O}\left(\frac{1}{z}\right)  \ .
\end{equation}
Hence this can be treated as an effective $4$ vertex. At this order in $z$  there is also a direct contribution of the usual Yang-Mills four vertex. These two terms can be summed to read in the large $z$ limit in color ordered form
\begin{equation}\label{eq:treelevellargez}
\epsilon_{1}^{\mu}\left(\hat{k}_1\right) \epsilon_{2}^{\nu}\left(\hat{k}_2\right) V^4_{\mu\nu \rho \sigma} \sim \epsilon_{1}^{\mu}\left(\hat{k}_1\right) \epsilon_{2}^{\nu}\left(\hat{k}_2\right) \left(\eta_{\mu \rho} \eta_{\nu \sigma} - \eta_{\nu \rho} \eta_{\mu \sigma}\right) \ ,
\end{equation}
where $\sigma$ and $\rho$ connect to the rest of the diagram. Note that this is anti-symmetric in the indices $\mu$ and $\nu$ (as well as $\mu$ and $\sigma$), although the two four vertex terms separately are not anti-symmetric\footnote{It is often said that BCFW recursion relations make the four point vertex of the Yang-Mills Lagrangian obsolete. While strictly true, the computation here shows clearly that in \emph{deriving} the relations the four vertex is crucial.}. 

Summarizing, in the limit of large BCFW shift of two color adjacent gluons the scattering amplitude scales as  
\begin{equation}\label{eq:largezform}
A(z) \rightarrow \epsilon_{1}^{\mu}\left(\hat{k}_1\right) \epsilon_{2}^{\nu}\left(\hat{k}_2\right) M_{\mu\nu}  = 
\epsilon_{1}^{\mu}\left(z \right) \epsilon_{2}^{\nu}\left(z\right) \left(z \, \eta_{\mu\nu} f_1(1/z) + f_{2,\mu\nu}\left(1/z\right) + \mathcal{O}\left(\frac{1}{z}\right) \right) \ ,
\end{equation}
with $f_1$ given in equation \eqref{eq:f1treelevel} and the antisymmetric term $f_2$ given by
\begin{equation}\label{eq:f2treelevel}
f_{2,\mu\nu} =  \left[(\eta_{\mu \kappa} (k_1+k_2)_{\nu} - \eta_{\nu \kappa} (k_1+k_2)_{\mu}) J^{\kappa}_q +  \sum_i \left(\eta_{\mu \rho} \eta_{\nu \sigma} - \eta_{\nu \rho} \eta_{\mu \sigma}\right)J_q^{\rho}(X_i) J_q^{\sigma}(X_{i+1})\right] \ ,
\end{equation}
where $J_q$ are again the currents calculated in AHK gauge and the sum runs over the ways to subdivide the set $X$ of external particles into two subsets $X_i, X_{i+1}$, keeping the color order of the particles. The structure displayed in equation \eqref{eq:largezform}  is equivalent to the result of \cite{ArkaniHamed:2008yf} but bypasses the use of the tree level field equations. Using the explicit scaling of the external wave functions reviewed in the appendix the analysis just presented leads directly to table \ref{tab:largezn4}.

\subsubsection*{Comparison to Feynman-'t Hooft gauge}
The above large $z$ contributions although obtained in a particular gauge, are expected to be gauge invariant since are the limit of a gauge invariant quantity. As a cross-check the leading order in $z$ part of the above result also follows from power-counting the diagrams which contribute at this order in Feynman-'t Hooft gauge. What changes here is the form of the propagator as well as the conditions on the fields connecting to the hard line. The leading $z$ contribution for Yang-Mills theory in this gauge is again formed by a hard line which only consists of three vertices. By the general structure of the three vertex the hard line will have two fields external to the hard line contracted with a metric, while the other fields are contracted into $q_{\mu}$: allowing other fields will lower the power of $z$. Therefore, for pure gauge theory the hard line contributes
\begin{equation}\label{eq:hardgluonline}
\sim [z] A^{1,\mu} \left(\eta_{\mu\nu} + K_{\mu} q_{\nu} + q_{\mu} \tilde{K}_{\nu} + f_1 q_{\mu} q_{\nu}  \right) A^{2, \nu} \ ,
\end{equation}
in the large $z$ limit, where $A^1_{\mu}$  and $A^2_{\nu}$ connect to the external wave functions. The quantities $K$ and $\tilde{K}$ are arbitrary vectors which stand for further connections in the diagram. From equation \eqref{eq:simplcontractwithq} it follows that the leading contribution from any diagram with a hard line is proportional to the metric contraction in pure Yang-Mills, while the proportionality factor is calculated in the same way as derived above in AHK gauge by contracting with the current. The powercounting argument in Feynman-'t Hooft gauge could probably be extended to the subleading contributions at order $z^0$ with more effort. 

\subsection{Adding scalar and quark matter}
For theories with more general matter content than tree level Yang-Mills matter couplings are important. The effects of these on shifts of gluons can be analyzed by a simple extension of the above.

\subsubsection*{Scalars}
Scalar matter will be important since ghosts for instance generically have scalar type couplings to glue. With the gluons in AHK gauge there is a limited number of new diagrams appearing for the BCFW shift of two gluons if coupling to scalar matter is included. These involve one or more scalar hard lines. For every connected scalar hard line it is clear that this part of the hard line will contribute to leading order
\begin{equation}\label{eq:scalarcurrenttohardline}
\sim A_{\mu} A_{\nu} \left(z \, q^{\mu} q^{\nu} + K^{\mu} q^{\nu}  + \tilde{K}^{\nu} q^{\mu} + \mathcal{O}\left(\frac{1}{z} \right)\right) \ ,
\end{equation}
where the explicit gluon fields are simply placeholders for any connection further along the hard line and $K$ and $\tilde{K}$ are again arbitrary vectors which stand for further connections in the diagram. In AHK gauge it should be noted that these vectors would always involve  either a $q^{\mu}$ connecting to a soft or to a hard line. For soft lines these diagrams actually vanish while for hard lines this contribution is suppressed by $\left(\frac{1}{z} \right)$. From the analysis of the gluon diagrams in AHK gauge it is clear that only those diagrams where the scalar line contains just zero or one element and where the outside legs connect directly to the scalar line will potentially contribute to the large shift limit. The diagrams with one element would always contain at least one of
\begin{equation}
k^1_{\mu} \epsilon_{1}^{\mu} \quad \textrm{or} \quad  k^2_{\mu} \epsilon_{2}^{\mu} \ ,
\end{equation}
as terms which have a possible z dependence in the denominator, but these terms vanish of course. The only remaining contribution is formed by the two hard lines connecting directly to the $2$ gluon $2$ scalar vertex. This is of course proportional to the metric and of order $\left(z^0\right)$. The effects of this vertex should be included in equation \eqref{eq:largezform} if one wants to calculate the exact coefficients of the large shift, with of course appropriate scalar currents attached. 

With the gluons in Feynman-'t Hooft gauge instead of AHK gauge equation \eqref{eq:scalarcurrenttohardline} continues to hold. This does not change the form of the current in equation \eqref{eq:hardgluonline} when scalars are included and hence therefore does not change the form of \eqref{eq:largezform}. 

\subsubsection*{Quarks}
Fermionic spin one half contributions to the hard gluon line are suppressed by one power of $z$ compared to the leading contribution in the gluonic case. This follows because the powers of  $z$ in the numerator are generated by the fermionic propagators, not by the vertices. In the case of minimally coupled spin one half matter, there are only three particle couplings. These lead for the diagrams which form a fermionic path along the hard line to
\begin{equation}\label{eq:fermcurrenttohardline}
\sim A_{\mu} A_{\nu} \overline{\psi}\left(\gamma_{\mu}  G \gamma_{\nu} + \mathcal{O}\left(\frac{1}{z} \right)\right) \xi \ ,
\end{equation}
where $\psi$ and $\xi$ are arbitrary spinors which symbolize the coupling to the soft parts of the diagrams and the matrix G reads
\begin{equation} 
G = \left(q \!\!\! \slash \right) \prod_{j=0}^{l} \left( \gamma_{\mu_j} \left(q \!\!\! \slash \right) \right) \ ,
\end{equation}
where the indices $\mu_j$ connect to soft parts of the diagram and there are $l$ external soft fields attached to the fermion line. Since $\left(q \!\!\! \slash \right)^2 =0$ one can also write this as
\begin{equation}
G = \left(q \!\!\! \slash \right) \prod_j \left( q_{\mu_j} \right) \ .
\end{equation}
The resulting gamma matrix structure can easily be split into a symmetric and anti-symmetric part to give
\begin{equation}
\sim A_{\mu} A_{\nu} \overline{\psi}\left(\gamma^{[\mu}  G \gamma^{\nu]} + \eta^{\mu\nu} G + \left( q^{\mu} \gamma^{\nu} + \gamma^{\mu} q^{\nu}    \prod_j \left( q_{\mu_j} \right)\right) + \mathcal{O}\left(\frac{1}{z} \right)\right) \xi \ ,
\end{equation}
where square brackets denote antisymmetrisation. In AHK gauge any non-trivial $q_{\mu_i}$ contract to the soft fields and the diagrams vanish, leaving the contribution with a single element. Just as above it is easy to see the hard fields have to attach directly to the fermion line with just one propagator to have a non-trivial contribution. This allows to drop the second symmetric term in the above as this is subleading in $z$ after contraction with the external hard wave functions. The remaining two terms  must be added to \eqref{eq:largezform} to calculate exact limits. 
 
With the gluons in Feynman-'t Hooft gauge it is easy to see that just as for scalars spin one half matter doesn't change the overall structure of the leading large $z$ limit on two gluons.

\subsection{Remarks}
Adding scalar fermion couplings does not change the preceding analysis as this coupling is automatically suppressed. It should be stressed that the derivation of the leading $z$ behavior as presented in this section applies to basically any minimally coupled gauge theory with scalar or fermionic matter. Also the above clearly shows that at leading order it is a property of a subset of the full Feynman graphs: only the hard line needs to be considered with a simplified set of Feynman rules. This shows that on-shell recursion techniques can be applied to more general correlation functions than those used for scattering amplitudes only. Also, since the leading large $z$ behavior arises from a either a three vertex diagram with the shifted legs attached directly or from a diagram with the shifted legs connected by one propagator only, it is easy to see that non-adjacent shifts are suppressed by at least an additional $\left(\frac{1}{z}\right)$.

\subsubsection*{Bonus relations?}
Note that knowledge of the exact form of the large BCFW shift limit of the scattering amplitudes can in principle be used to derive new relations between scattering amplitudes at tree level. This follows because the BCFW derivation of equation \eqref{eq:BCFWrecursionint} can now be modified as
\begin{equation}
0= \oint_{z=0} (1 + \alpha z) A(z) = \sum_{res} \widetilde{\left(z = \textrm{finite}\right)} +  \sum_{res}  \widetilde{\left(z = \textrm{infinite}\right)} \ ,
\end{equation}
for some arbitrary coefficient $\alpha$. Here the modified finite $z$ residues are now a sum over products of tree amplitudes times a momentum dependent function. This sum equals the now non-vanishing residue at infinity which can be calculated explicitly for the $(+-)$ good shift through the above diagrammatic techniques for instance. For shifts of a pair of helicity equal gluons there is a gap consisting of the missing explicit order $\left(\frac{1}{z} \right)$ contributions from the diagrammatic analysis for this particular shift. It would be interesting to explore this further as the currents can be calculated using auxiliary recursion relations if needed. 
 
\subsubsection*{Currents}
In the above one leg off-shell currents appear, calculated in two different gauges. However, it is fairly easy to see that the currents themselves are largely gauge invariant as they can be calculated through an auxiliary BCFW recursion relation. For this one simply uses an auxiliary BCFW shift. The resulting recursion relations, when iterated, only rely on a new three vertex: the three particle current with one leg off-shell. Since for this vertex 
\begin{equation}
(K)^{\mu} J^{3}_{\mu} =0 
\end{equation} 
holds with $K$ the momentum in the off-shell leg, it holds for the $n$-particle current as well. Using off-shell recursion relations this was proven in \cite{berendsgiele}. Note also that the current is invariant under gauge transformations on each leg separately, i.e. the current is invariant under
\begin{equation}
\xi_i \rightarrow \xi_i + k_i \ .
\end{equation}
The same conclusion follows by a short modification of the argument above for the singular diagrams in the AHK gauge: requiring absence of gauge irregularities in the $x \rightarrow 0$ limit shows that the current must be gauge invariant if the parent amplitude is.
 
\section{BCFW shifts of loop amplitude integrands}
\label{sec:bcfwloopsintegr}
In this section the discussion is extended to loop level. There are some obstructions to applying the above tree level derivation of on-shell recursion relations and in particular of equation \eqref{eq:BCFWrecursionint} to the loop level directly. Most obviously, at loop level there will be branch cuts in the $z$ plane in general such as those sketched in figure \ref{fig:BCFWloops}. 

\FIGURE[h]{\epsfig{file=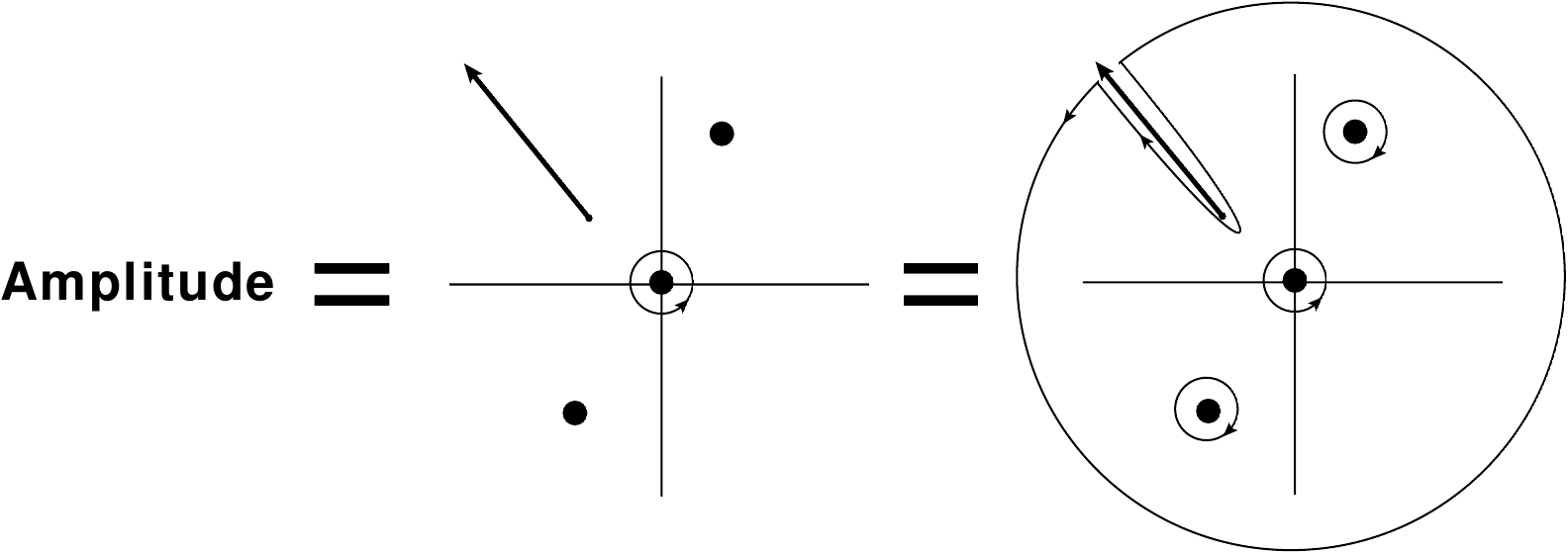,scale=0.7} \caption{Sketch of the contours used for BCFW recursion at loop level } \label{fig:BCFWloops}}

Note however that the contour integral at $z=0$ can still be deformed to encircle the branch cuts. Assuming the branch cuts do not overlap, one can see that the integral around the branch cuts is simply an integral over the imaginary part of the discontinuity across the branch cut. Since the discontinuities in the branch cuts in amplitudes generically correspond to cut propagators in the Feynman graphs, one sees immediately that there is scope for a loop level analogue of on-shell recursion. For this to work properly, the integrals over the arc at infinity must still vanish or at least give a well-defined result. Hence for immediate applications one still would like the amplitude to scale well at $z=\infty$, as long as $z$ is not directed along one of the branch cuts.

Sidestepping integrated amplitudes first, in this section the previous discussion is extended to loop level integrands. This is natural as the integrand is just as the tree level amplitude a rational function of the momenta. Hence it scales as a rational function of $z$. The main argument is based on powercounting, but the main idea of the present section first arose from a simple consideration in string theory which will be presented first. Readers not interested in string theory should probably skip the next subsection.  

\subsection{String theory perspective}
 \label{subsec:stringmot}
On-shell recursion at string tree level has recently been proven using a certain limit of the OPE between two adjacent vertex operators \cite{Cheung:2010vn} \cite{Boels:2010bv}.  This argument can be applied to a general string theory amplitude which reads for the closed string for instance
\begin{equation}
A_n = \int d\mu_l \langle V_{1}(z_1) \ldots V_{n}(z_n)\rangle \ ,
\end{equation}
where $d\mu$ is the measure on the moduli space of spheres with $l$ handles and $n$ operator insertions. Despite its complicated form the moduli space integral does not depend on the momenta of the particles.  

On a genus $g$ Riemann surface the OPE between for instance two closed string tachyon vertex operators in the bosonic string reads
\begin{equation}\label{eq:OPEexamp}
:\!V(z_1) \!:  :\!V(z_2) \!: \sim  e^{G_g  (k_1 \cdot k_2 )(z_1-z_2)}:\! V(z_1) V(z_2)  \!: \ ,
\end{equation}
where $G_g$ is the solution to the equation
\begin{equation}
\partial_w \partial_{\bar{w}} G_g(w) = \delta^2(w) 
\end{equation}
on the Riemann surface of interest. Crucially, for small $w$ $G_g$ tends to the tree level two point correlator on the sphere,
\begin{equation}
G_g(w) \sim \log(w) + \log(\overline{w}) + {\mathcal{O}(w)} \ .
\end{equation}
It is easy to see that the prefactor on the right hand side in equation \eqref{eq:OPEexamp} is shift independent because it involves $k_1 \cdot k_2$. In other words, if we assume just as at tree level that $wz \sim 1$ and take the large $z$ limit, the analysis quickly reduces to the analysis of BCFW at string tree level discussed in \cite{Cheung:2010vn} and \cite{Boels:2010bv}. For more general shifted particles in both open and closed (super)strings a similar derivation also holds. Ignoring the measure and possible complications from singular points on the moduli space, the above suggests the natural extension of suspicion \ref{sus:main} to string theory:

 \begin{suspicion}\label{sus:string}
In four dimensions and up, away from singular points on the moduli space, the integrand of open or closed string theory amplitudes in a flat background scales under a BCFW shift of a pair of outside legs (at least as good) as the tree string amplitude to all loop orders. 
\end{suspicion}

Taking the naive field theory limit of this string theory suspicion gives the above field theory suspicion (\ref{sus:main}). This explicitly ignores any possible complications with the decoupling of massive modes for instance. Note that the above argument holds for adjacent vertex operators only in the case of the open string. Non-adjacent shifts would be expected to be exponentially suppressed. 

\subsection{Shifts of the integrand at arbitrary loop level}
Note that even before beginning it is expected that the integrand of loop level amplitudes will scale as a rational function of $z$. This simply follows from any representation of the integrand of the  loop amplitude in terms of Feynman diagrams. The large $z$ shifts will involve shifts of the loop propagators by
\begin{equation}
\frac{1}{(l_i+k_1+K)^2 + m_i^2} \rightarrow \frac{1}{(l_i+k_1+ K)^2 + 2 z q \cdot (l_i+K) + m_i^2} \ , 
\end{equation}
for a shift on momentum $k_1$ within a loop made of a particle of mass $m_i$ and loop momentum $l_i$, where $K$ is additional unshifted momentum (which explicitly rules out $k_2$). As long as the loop momentum and the external momenta are real and generic
\begin{equation}
q \cdot (l_i +K) \neq 0 
\end{equation}
will hold and naive powercounting can be applied. Whenever the counting of $z$ factors employed below will fail is where
\begin{equation}\label{eq:restronqloops}
 q \cdot (l_i+K)  = 0 \ , 
\end{equation}
or where the loop momentum is infinite. For real momenta the condition \eqref{eq:restronqloops}  defines a $D-2$ dimensional sub-manifold for the loop integration. As long as the integrand does not have delta-function support on this manifold (which it will not have for generic and real external momenta), these regions will not contribute to the amplitude. More generally, one can study points where the loop momentum is also sufficiently generic. This allows one to bypass the reality constriction on the loop momenta. 

For the argument below to be transparent it is convenient to choose a convention for how the momenta in the loop depend on the external momenta: after all this can be changed by a linear, finite shift of any of the loop momenta. This shifts the integrand by a function which integrates to zero. For convenience one can choose the convention for the routing of the loop momenta such that the $z$-dependence in the loop diagram flows along a minimal path between the legs with shifted momenta, i.e. along the hard line. That such a routing exists for shifts of two particles within the same color trace can be seen by following the path on the color trace: the routing condition determines a convention for each loop encountered along the path. This fixes the $z$ dependence of all loop momenta, leaving non-$z$-dependent loop momenta undetermined. 

For shifts of particles on different color traces no canonical choice of hard line exists. At one loop shifts on different color traces can be related to shifts of particles on the same color trace through the relations derived in \cite{Bern:1994zx}. Hence this is only an issue at two loops and beyond this case will not be studied in this article, save for the formulation of the suspicion that shifts of particles on different color traces are suppressed by at least one power of $1/z$ compared to a shift of color adjacent particles. This depends on the ability to choose the loop momentum dependence on $z$ to be only along a particular line in the diagram: following through the diagrammatic analysis this yields an $1/z$ suppression. Again, this case will not be further considered here.

\subsubsection*{Powercounting}
The crux of the argument in this section is that the above diagrammatic analysis of the shift behavior in AHK gauge although motivated at tree level did not depend on on-shell constraints on the soft external legs at any point: it applies to a wide class of correlation functions. In particular, it does not depend on the loop order in an amplitude calculation. Up to a minor subtlety discussed below this will lead to the conclusion that the same analysis presented holds for the sum over diagrams in this particular gauge at any loop order, and in particular the integrand shifts according to the general structure outlined in equation \eqref{eq:largezform}. Here the functions $f_1$ and $f_2$ in this Ansatz will not be given by \eqref{eq:f1treelevel} and \eqref{eq:f2treelevel} any more but by suitable generalizations. Hence the integrand of gauge theories coupled to various forms of matter scales as in table \ref{tab:largezn4} to all loop orders for shifts of color adjacent gluons. The analysis of shifts of fermions considered in the appendix also generalizes immediately. 

There is a subtlety in the above reasoning which has to do with singular diagrams in the AHK gauge: there is an additional class of diagrams which consist of (generalized) 'triangle' shaped loop diagrams on which the shifted legs end directly and only one leg connects to the rest of the diagrams. Just as at tree level this is singular in AHK gauge because the momentum $k_1 + k_2$ flows through the off-shell leg. In fact, this class has two different elements.  One is formed by coupling the hard fields to the effective four vertex and tying the remaining two legs at the required loop order into a single external field. This diagram is of order $\left(z^0\right)$ and either proportional to the metric or anti-symmetric by the structure of the effective four vertex.

The other is formed by the tree level three vertex coupling to the hard fields and with a propagator correction graph on the external line. This has the same form as the tree level contribution (order $(z^1)$ and $(z^0)$). This can be seen in a bit more detail. Since the propagator correction has to be transversal, the denominator of this correction must have the tensor structure
\begin{equation}
\eta^{\mu\nu} k^2  - (k_1+k_2)^{\mu} (k_1+k_2)^{\nu} \ , 
\end{equation}
to any loop order. Contracting with two propagators in the modified AHK gauge as above gives simply a tensor structure of the lightcone gauge propagator plus an extra piece of the form
\begin{equation}
\sim \frac{(q+ x k_1)^{\mu} (q+ x k_1)^{\nu} }{(x (k_1 + k_2))^2} \ .
\end{equation}
This extra term does not contribute since it contracts into the soft part of the diagrams. This reduces the analysis to terms studied before at tree level. 

Note that calculating loop integrals in a lightcone gauge directly can be fraught with difficulties as poles of the form
\begin{equation}
\sim  \frac{1}{q\cdot l_i} \ ,
\end{equation}
arise which can however be overcome, see \cite{Mandelstam:1982cb} and \cite{Leibbrandt:1983pj}. These poles cancel within the full amplitude expression as they should by gauge invariance, but can be hard to treat directly. 

In Feynman-'t Hooft gauge the leading order $\left( z^1\right)$ behavior of the integrand can also be reproduced following the same reasoning as at tree level. Similarly, in this gauge the sub-leading behavior is harder to see. 

\subsection{Remarks}
There are a number of natural expectations which follow from the above result. In brief, for every theory for which the BCFW shift behavior has been analyzed at tree level one expects the same behavior for the loop level integrand, to all loop orders. In $\phi^4$ theory for instance this is easy to verify as the on-shell conditions for the external particles have no impact on the powercounting. For Einstein gravity the same is expected to hold, but this is although much more interesting also much harder to prove. As touched upon above as motivation, it is expected that for string theory a similar phenomenon holds for the integrand. 

Most of the explicit results above are for color-adjacent gluons. Note though that the argument does not depend on the color details of the rest of the diagrams: it holds for shifts of any two color adjacent gluons. In particular it holds for non-planar diagrams. Further, one can generalize easily the above to shifts of gluons on the same color trace as there still is a canonical hard line along the edge of the diagram. As discussed before, shifts of gluons on different color traces beyond two loops will be left to future work.

A natural extension of the results of this section is to use coherent state methods (BCFW `supershifts') in maximally supersymmetric theories in four dimensions \cite{Brandhuber:2008pf}, \cite{ArkaniHamed:2008gz} or higher \cite{Boels:2009bv}. This can also be used for instance to derive a slate of predictions for shifts of any pair of members of the $N=4$ multiplet, at any point of the moduli space \cite{Boels:2010mj}.  

\section{Difference between integrand and integral at one loop}
\label{sec:integralsvsintegrands}
After the discussion of the BCFW scaling of the integrand it is a natural question how this extends to the scaling of the integrals of these integrands. After all, one would like to study recursion for the amplitudes directly as these are the physical objects. The difference between the two scalings depends on the ability to interchange integration and limit. This can be phrased as an order of limits problem: the loop integrals are regularized to make them well-defined by, say, dimensional regularization. Then the question is whether the limit of large BCFW shift ($z \rightarrow \infty$) commutes with the limit in which the dimensional regularization parameter vanishes ($\epsilon \rightarrow 0$). For the leading singularities of the integrands this is (almost) bound to work from generic expectations of UV and IR divergences. Wether this also holds for the sub-leading in $\epsilon$ parts of the calculation is a question which will be studied in this section. 

In general one should be looking to ways in which the powercounting argument presented above could fail after integration. By the general argument just mentioned it is natural to suspect the UV and IR divergences of the integrals. The first arise from the momentum integration directly, while the second are related to the Feynman parameter integrals. To get a grip on the situation and also to be able to compare to results in the literature most of the concrete results in this section will be obtained at one loop.

Another way of phrasing the problem under study is the question whether amplitudes are `single-cut' constructible. As outlined above, the singularities of the integrand should be related to the branch cuts in the $z$ plane. Any possible discrepancy arises from the integral over the arc at infinity. Since this is technically much easier, one would like to calculate cuts in four dimensions, not $(4-2\epsilon)$ of them. This gives information on the leading poles of the $\epsilon$ expansion, but disregards information about sub-leading terms. The question is if the information of the leading poles is enough to determine the full amplitude. In other words, the question is if boundary terms are missed if the limit $\epsilon \rightarrow 0$ is taken before the $z \rightarrow \infty$ one. Based on known results \cite{Bern:1994zx} \cite{Bern:2005hs} it is easy to speculate that this possible problem is absent in supersymmetric theories with massless matter but is present in non-supersymmetric gauge theory. 

In the following we are in particular interested to find deviations from the general form of the tree level amplitude at large $z$ as given in equation \eqref{eq:largezform} as this would signal a key difference between integral and integrand. 

\subsection{BCFW shift of the integrals at one loop}
In this subsection the powercounting argument will be applied to the (gluonic sector of the) one-loop quantum effective action as calculated in the background field method \cite{Abbott:1981ke} \cite{Abbott:1983zw}. The quantum fields will be put in background Feynman-'t Hooft gauge, while the tree level fields will be kept in AHK gauge. One direct motivation to study background field methods is that in this gauge in Yang-Mills theory it is easy to see supersymmetric cancellations at the one loop level directly. Moreover, it allows us to disregard the problems of the integration of unphysical poles in the momentum integration associated to the lightcone gauge, while keeping the analysis of tree level diagrams presented above intact. 

The gauge invariant one-loop contributions to the effective action as calculated through the background field method read
\begin{equation}\label{eq:1loopcontritoeffac}
\mathcal{L}_{\textrm{eff}} = \mathcal{L}_{\textrm{tree}} + \mathcal{L}_{\textrm{scalars}} + \mathcal{L}_{\textrm{fermions}} + \mathcal{L}_{\textrm{gluon}} \ ,
\end{equation}
with
\begin{align}
\mathcal{L}_{\textrm{scalar}} & = \log \det\!\!\!\phantom{|}^{-1}_{s=0} \left(D_{\mu} D^{\mu} \right) \\
\mathcal{L}_{\textrm{chiral fermion}} & =  \log \det\!\!\!\phantom{|}^{\frac{1}{2}}_{s=\frac{1}{2}} \left(D_{\mu} D^{\mu} + \sigma_{\mu\nu} F^{\mu\nu} \right) \\
\mathcal{L}_{\textrm{gluon}} & = \log \det\!\!\!\phantom{|}^{-\frac{1}{2}}_{s=1} \left(D_{\mu} D^{\mu} +  \Sigma_{\mu\nu} F^{\mu\nu} \right) + \log \det\!\!\!\phantom{|}^1_{s=0} \left(D_{\mu} D^{\mu} \right) \ .
\end{align}
Here $D_{\mu}$ is the covariant derivative for the external field $A$, $\sigma$ is the Lorentz generator in the spinor representation ($\sigma_{\mu\nu} = \frac{1}{4} [\gamma_\mu, \gamma_\nu]$) and $\Sigma$ is the Lorentz generator in the vector representation. The determinants are shorthand for the path integrals to be calculated which generate the Feynman graphs for the vertices. Explicit color-ordered Feynman rules can be found for instance in figure (8) in \cite{Bern:1996je}. Explicit results for the three and four vertex as calculated through the background field method obtained with a different motivation can be found in \cite{Denner:1994nn},\cite{Hashimoto:1994ct}. 

In the Feynman-'t Hooft background field gauge the vertices of the quantum effective action are all proportional to integrals of the form 
\begin{equation}\label{eq:genpowercountoneloop}
I^k_{n} = \int \frac{d^D l}{(2 \pi)^D} \frac{l^{\mu_1} \ldots l^{\mu_k}}{l^2 (l+k_1)^2 \ldots (l+\sum_{i=1}^{n-1} k_i)} \ ,
\end{equation}
where $k_i$ are the external momenta and there are $k$ powers of $l$ in the numerator. By convention the shifted momentum will be taken to enter the color ordered loop loop at the leg with momentum $k_1$ and to exit at the leg with momentum $k_2$ as the focus will primarily be on shifts of color-adjacent graphs. The generalization to more `distance' between the shifted legs is trivial for the general argument to be presented below. $D$ denotes the space-time dimension of the integrals, which will be taken to be $D = 4-2 \epsilon$ below\footnote{We will opt to keep Lorentz indices in the numerator in $D$ dimensions as this is more convenient in an off-shell formalism.}. In gauge theory $k\leq n$ holds by simple powercounting. Incidentally, note that the most UV divergent graph ($k=n$) consists of three point vertices only which is also the class of diagrams important to the large BCFW shift limit in Feynman-'t Hooft gauge at tree level. In the following some details will be needed about the structure of the integrations appearing at the one loop level. 

\subsubsection*{Momentum integrals}
The standard method of calculating the class of integrals in equation \eqref{eq:genpowercountoneloop} proceeds through the introduction of Feynman parameters 
\begin{equation}
I^k_{n} = \Gamma(n) \int da_n  \int \frac{d^D l}{(2 \pi)^D}  \frac{l^{\mu_1} \ldots l^{\mu_k}}{(\alpha_1 l^2 + \alpha_2 (l+k_1)^2 +  \ldots + \alpha_n (l+\sum_{i=1}^{n-1} k_i))^n} \ ,
\end{equation}
with
\begin{equation}
 \int da_n   \equiv  \int_0^1 \left(\prod_{i=1}^{n} da_i\right)   \delta(\sum_{i=1}^n \alpha_i -1) \ ,
\end{equation}
followed by a shift on the integration variable,
\begin{equation}
l^{\mu} \rightarrow l^{\mu} + \delta^{\mu} \ ,
\end{equation}
with
\begin{equation} \label{eq:exprfordelta}
\delta^{\mu} = - \sum_{i=2}^{n} \left(\sum_{j=1}^{i-1} k_j \right) \alpha_i \ ,
\end{equation}
which brings the integral into the form
\begin{equation}
I^k_{n} = \Gamma(n) \int da_n  \int \frac{d^D l}{(2 \pi)^D}  \frac{(l+\delta)^{\mu_1} \ldots (l+\delta)^{\mu_k}}{(l^2 + \Delta)^n} \ .
\end{equation}
In this expression
\begin{equation}\label{eq:exprforDelta}
\Delta = \sum_{i=1}^n \alpha_i(\sum_{j=0}^{i-1} k_j + \delta)^2 \ ,
\end{equation}
where $k_0 =0 $ is defined for convenience. At this point the momentum integral can be performed. This follows from the repeated use of
\begin{equation}
l^{\mu_i} l^{\mu_j} \rightarrow l^2 \frac{\eta^{\mu_i \mu_j}}{D}  \ ,
\end{equation}
which follows from spherical symmetry. This leads to the generic momentum integral formula
\begin{equation}
\int \frac{d^D l}{(2 \pi)^D}  \frac{l^{\mu_1} \ldots l^{\mu_{2i}}}{(l^2 + \Delta)^n}= \frac{(-1)^n i}{(4 \pi)^{D/2}} \frac{ \Gamma(n-\frac{D}{2} -i)}{2^i \Gamma(n)}  \eta_s^{\mu_1 \ldots \mu_{2i}}  \left(\frac{1}{\Delta}\right)^{n-\frac{D}{2} -j} \ ,
\end{equation} 
where $\eta_s^{\mu_1 \ldots \mu_{2i}}$ is a sum over products of metrices with completely symmetrized indices. This leaves a generically complicated integration over the Feynman parameters.

\subsubsection*{Feynman parameter integrals}
The integration over Feynman parameters is generically of the form
\begin{equation}\label{eq:genfeynparamintegr}
I \sim \int da_n \frac{f(\alpha_i)}{\left(\sum_{i,j=1}^n S_{ij} \alpha_i \alpha_j\right)^{n-D/2}}  \ ,
\end{equation}
with $f$ a polynomial of the Feynman parameters of maximal degree $n$ and $S_{ij}$ is some symmetric matrix which depends on external momentum invariants whose form is unimportant at the moment. There is a special sub-class of these integrals which can be integrated easily, 
\begin{equation}\label{eq:feynparamintegr}
 \int da_n \prod_{i=1}^k \left(\alpha_i\right)^{\beta_i} = \frac{1}{\Gamma(k + \sum_{j} \beta_j)} \prod_{i=1}^{k} \left(\Gamma(1 + \beta_i)\right) \ ,
\end{equation}
which holds as long as the arguments of the gamma functions all have positive real part,
\begin{equation}\label{eq:realityconstraintsmellbarn}
\Re (1+\beta_i) >0 \qquad \forall i \ .
\end{equation}
At this point one can use the general approach of Mellin-Barnes integration (see e.g. \cite{Smirnov:2006ry} and references therein) to reduce the starting Feynman parameter integral to the special sub-class. The general formula needed to make this work is
\begin{equation}\label{eq:mellinbarnes}
\frac{1}{(X+Y)^\lambda} = \frac{1}{\Gamma(\lambda)} \frac{1}{2 \pi i} \int_{- i \infty}^{+ i \infty}  dw \frac{X^w}{Y^{\lambda+w}} \Gamma(w+\lambda) \Gamma(-w) \ ,
\end{equation}
where the integral is along a contour from $- i \infty$ to $+ i \infty$ which splits the two series of poles of the Gamma function. This contour will be taken to be a straight line. By encircling either series of poles and summing the resulting residue integrals one can see that the above formula is closely related to Newton's binomial formula. Combining multiple applications of equation \eqref{eq:mellinbarnes} and the integration of \eqref{eq:feynparamintegr} one can see that one can trade the generic integral over Feynman parameters of equation \eqref{eq:genfeynparamintegr} for multiple Mellin-Barnes type integrals. 

One difficulty of the outlined approach is that in general the constraints of equation \eqref{eq:realityconstraintsmellbarn} may be violated for small values of $\epsilon$. One solution to this is to first continue $\epsilon$ to a value for which a good straight line contour can be found. Graphically, this puts the series of poles of all the involved $\Gamma$ functions to the left (for $\Gamma(x)$ terms) and right (for $\Gamma(-x)$ terms) of the chosen contour. Then $\epsilon$ is tuned back to a small value. This procedure will pick up residue integrals where poles cross the contour which can be calculated separately. This particular procedure is conveniently implemented in publicly available code \cite{Czakon:2005rk}.

\subsubsection*{Example: massless box}
As the Feynman parameter integrals may not be as familiar as the momentum integrals, perhaps an example is in order. For this, consider the Feynman parameter integral appearing in the calculation of the massless box, 
\begin{equation}
B = \int da_4 \frac{1}{\left( s \alpha_1 \alpha_3 + t \alpha_2 \alpha_4 \right)^{2 + \epsilon}}   \ .
\end{equation}
Following the outlined program above one arrives at the following Mellin-Barnes type integral
\begin{equation}
B = \frac{1}{t^{2 + \epsilon}} \frac{1}{\Gamma(-2 \epsilon)} \int_{- i \infty}^{+ i \infty} dw \left(\frac{s}{t}\right)^w \Gamma(w+2+\epsilon) \Gamma(-w) \Gamma(-w-1 - \epsilon)^2 \Gamma(w+1)^2   \ .
\end{equation}
To make this integral well-defined a contour needs to be specified. It is not too hard too see a straight line contour with $\Re(w) = - \frac{1}{2}$ for $\epsilon = -1$ will neatly divide the poles of the $6$ Gamma functions. Continuing $\epsilon$ to zero then picks up a residue integral at $w = -1-\epsilon$,
\begin{equation}\label{eq:telltalefeautreoflift}
B  \sim \left(\frac{1}{t s^{1 + \epsilon}} \right) f(\epsilon) + \left(\frac{1}{t^{2 + \epsilon}} \right) \int_{- i \infty}^{+ i \infty} R \ ,
\end{equation}
for some function $f$ of $\epsilon$ and a remaining Mellin-Barnes integral over a remaining integrand denoted by $R$. The function $f$ has a $\left(\frac{1}{\epsilon^2}\right)$ type pole typical of an IR divergence. Now suppose a BCFW shift of the box is considered of the type
\begin{equation}
s \rightarrow s \qquad t \rightarrow t + z' \ ,
\end{equation}
In this case it is seen that the integral scales one order of $z'$ better than the integrand.

\subsection{One loop momentum integrals contributing to large BCFW shift}
After this brief general discussion of the integrals appearing in the calculation of the one loop quantum effective action the next step is to discuss which one-loop diagrams contribute to the large BCFW limit of complete one loop amplitudes. For this one considers the one-loop vertices with hard lines entering on adjacent positions calculated using color ordered Feynman rules. In the following the focus will be on momentum integration, leaving the Feynman parameters for remarks at the end of the section.

A simple class of diagrams is if the hard line only 'glances' the loop (that is, it connects to the loop through one of the four-vertices with two background fields). The one-loop vertex is of order $\left(z^0\right)$ as it is either proportional to a metric or is anti-symmetric. From the tree level  one can simply use the tree level analysis to show that those diagrams where both shifted legs end on the loop vertex directly will not contribute new types of term to the generic large $z$-form of equation \eqref{eq:largezform}. A new class is formed by diagrams dressed by one order $z^0$ tree level hard leg as obtained from equation \eqref{eq:AHKgaugeprophardline}. For terms proportional to the metric this does not change the analysis, but for the anti-symmetric terms new order $z^0$ possibilities arise. It can be checked that this does not influence the $4$ point amplitude, but it is an issue for higher points. 

This leaves diagrams where there is at least one loop edge along the hard line which requires more care. First consider the effects of the momentum integration. One can see from equations \eqref{eq:exprfordelta} and \eqref{eq:exprforDelta} that in general both the shift $\delta^{\mu}$ and the function $\Delta$ are linear in $z$,
\begin{eqnarray}
\delta^{\mu} &\rightarrow & \delta^{\mu} - z a_2 q^{\mu} \\
\Delta &\rightarrow & \Delta -  2 z   a_2   \sum_{i=1}^n a_i(\sum_{j=0}^{i-1} q_j + \delta) \cdot q \label{eq:triangleshapedloophole} \ .
\end{eqnarray}
There is one loophole here where the two shifted external wave functions connect to the loop directly and form a triangle shaped diagram. The involved diagram topologies are illustrated in figure \ref{fig:1loopvertstriangles}. In this case the two external momenta appearing in $\Delta$ are both orthogonal to $q$ and hence $\Delta$ is in this case independent of $z$. Note that the same class of diagrams also contains those diagrams which are singular for the AHK gauge on the tree level parts as a sub-class. 

\FIGURE[!h]{\epsfig{file= 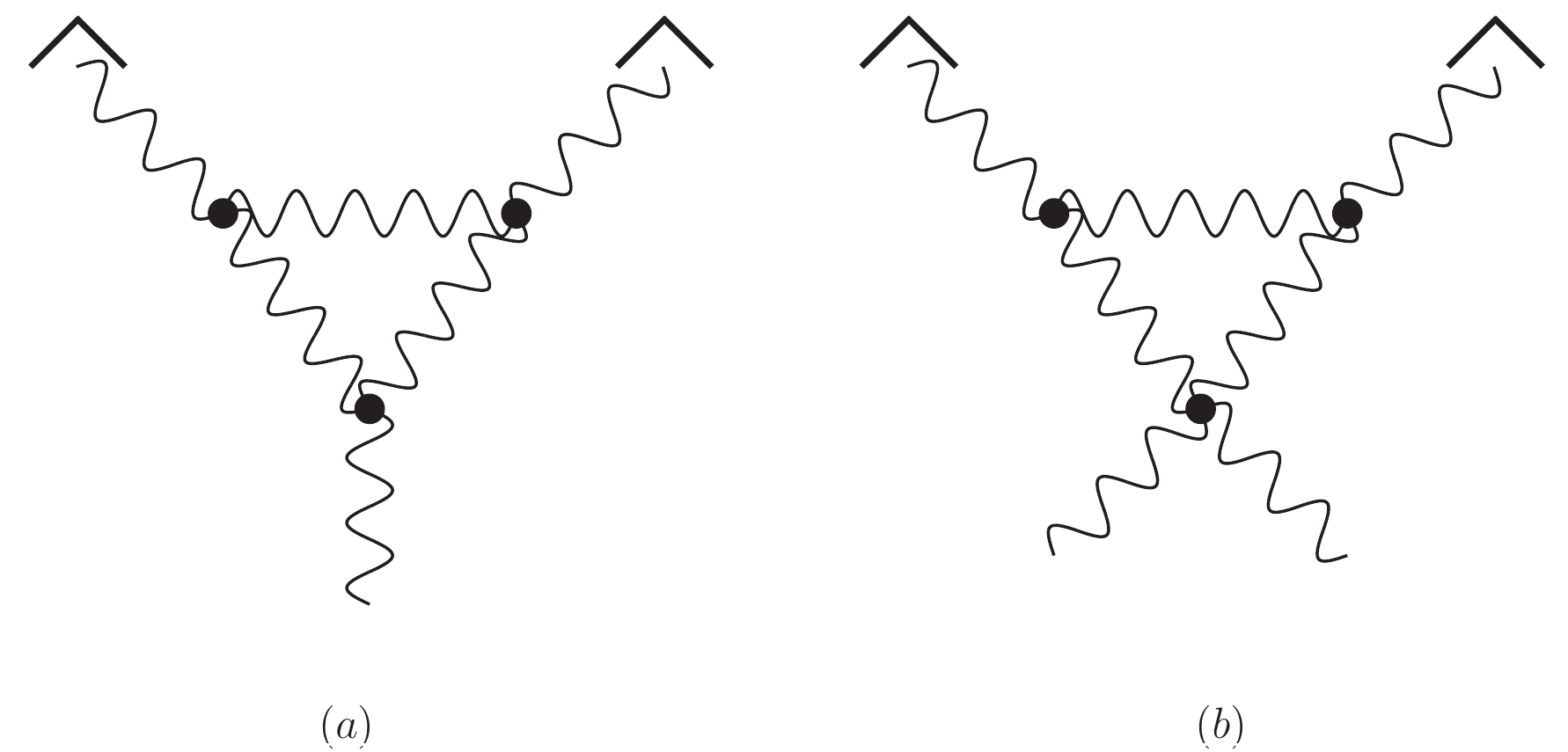,scale=0.5} \caption{diagrams corresponding to a triangle shaped loophole to the scaling of equation \eqref{eq:triangleshapedloophole}}  \label{fig:1loopvertstriangles}}

After momentum integration the $n$-point vertices in the quantum effective action at one loop where the shifted legs connect to different vertices on the loop integral are proportional to 
\begin{equation}
V_{n} \sim \sum_{j=0}^{[n/2]} c_{k,n} (z)^{D/2-n+j} \ ,
\end{equation}
for some field dependent coefficients $c_{k,n}$. Here $[n/2]$ is the greatest integer for which $[n/2] \leq n$. This allows us to classify the possible troublesome terms in $D=4-2 \epsilon$, which are at:
\begin{equation}
\begin{array}{ccc}
\quad n\quad & \quad k\quad & z-\textrm{order}\\
2 & 2 & z^{1-\epsilon} \\
2 & 0 & z^{0-\epsilon} \\
3 & 2 & z^{0-\epsilon} \\
3 & 3 & z^{0-\epsilon} \\
4 & 4 & z^{0-\epsilon} 
\end{array} \ ,
\end{equation}
while the three particle vertices have a maximal divergence degree of $z$. Using equation \eqref{eq:1loopcontritoeffac} and taking into account the possible tree level contributions this table can be translated into diagrams again. The origin of the diagrams from the two different types of vertices in the determinants (either the  $D^2$ or the $F$ type) will be stressed as this allows an easy way of comparing theories with different matter content. As a complication, the tree level AHK gauge propagator has a potential $z^0$ piece, see equation \eqref{eq:AHKgaugeprophardline}. This is proportional to two $q$'s though so this can only contribute in exceptional cases. In the figures below which correspond to the suspect diagrams this exceptional part of the AHK propagator along a hard line is indicated by a crossed out gluon propagator.

\FIGURE[!h]{\epsfig{file= 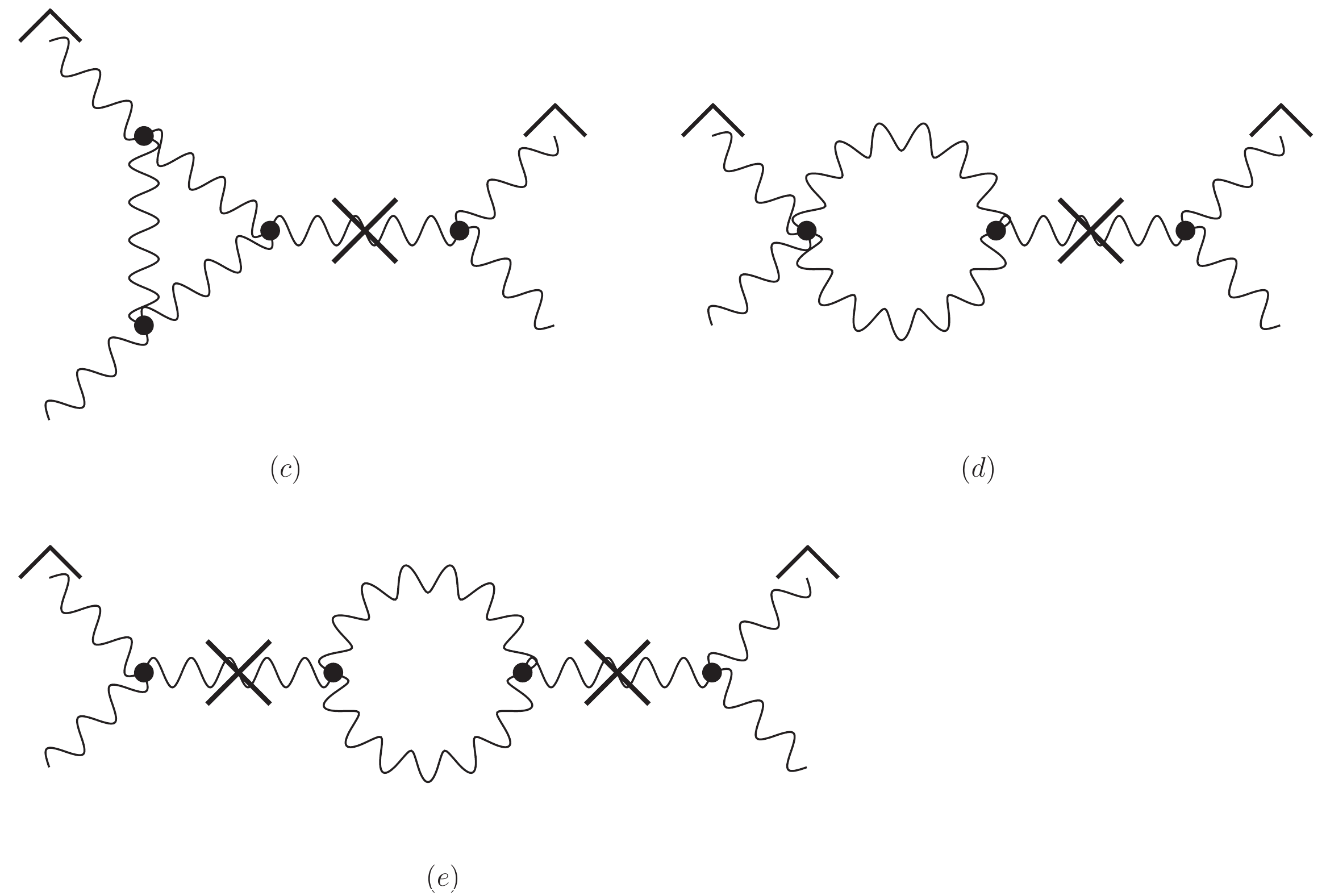,scale=0.5} \caption{Diagrams with one or two tree level legs along the hard line}\label{fig:1loopvertswlightconepieces}}
\FIGURE[!h]{\epsfig{file= 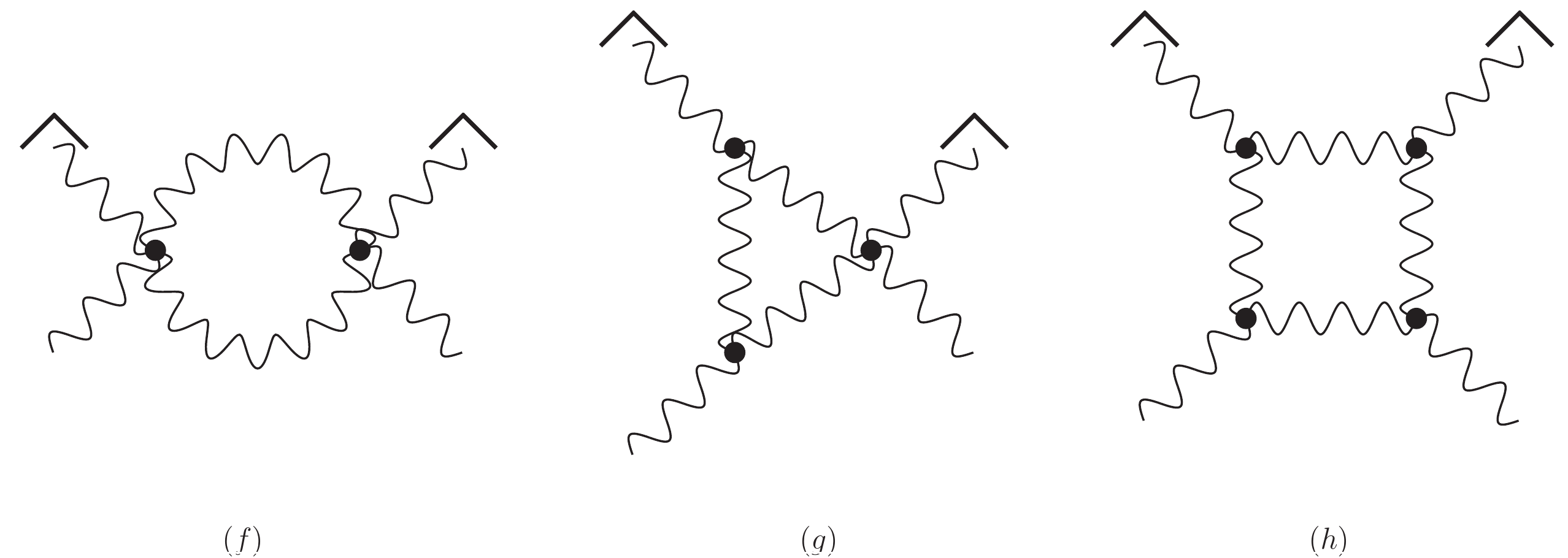,scale=0.5} \caption{Diagrams connecting the shifted legs directly to the vertex} \label{fig:1loopverts4pts}}

The listed Feynman graphs are calculated in terms of integrals in appendix \ref{app:loopintegrals}, up to terms which vanish in the large $z$ limit as analyzed through the momentum integrations. By tracing through the terms several results follow. The leading term in $z$ is given by diagram (a) on the left hand side of figure \eqref{fig:1loopvertstriangles} whose value is calculated in equation \eqref{eq:loopa}. This three particle diagram neatly parallels the diagram dominating at tree level. Structurally this term yields a contribution to the large $z$ behavior of
\begin{multline}
\epsilon^{\mu}_1 \epsilon_2^{\nu}  A^{(a)}_{\mu\nu\rho} \sim \epsilon^{\mu}_1 \epsilon_2^{\nu} \left(\frac{1}{2 k_1 \cdot k_2}\right)^{\epsilon}  \Gamma(\epsilon)  \left( [z] \left[\eta^{\mu\nu} f^1_{0} + \epsilon \frac{ k_{3,\mu} k_{3,\nu}}{k_3^2} f^{1}_1 \right] + f^1_{2, \mu\nu} [z^0] + \mathcal{O} \left(\frac{1}{z} \right) \right. \ .
 \end{multline}
Here $f_{2,\mu\nu}$ is antisymmetric and the functions $f_i$ have expansion in $\epsilon$ of the form
\begin{equation}
f_i \sim 1 + \mathcal{O} (\epsilon) \ , 
\end{equation}
from the Feynman parameter integrals. A $\frac{1}{\epsilon}$ divergence in $f_2$ arises from the Feynman parameter integral.  

The form of the exact large shift terms is found by contracting a tree level three vertex onto the one-loop three vertex. By the analysis of the gauge singularities there, the leading term in $z$ arises by contracting two tree level currents with a metric and multiplying this with
\begin{equation}
A(z) \sim \left(J_{\rho}(K_3) J_{\sigma} (K_4) \eta^{\rho\sigma} \right) \frac{\left(q \cdot K_4 \right)}{k_3^2} \epsilon^{\mu}_1 \epsilon_2^{\nu} \left[\eta_{\mu\nu} \Gamma(\epsilon) f^1_{0} + \epsilon  \frac{ k_{3,\mu} k_{3,\nu}}{k_3^2} f^{1}_1 \right] + \mathcal{O} \left(z^0 \right) \ ,
\end{equation}
where $K_4$ is the total momentum of all particles connecting to the current $J_{\sigma} (K_4)$. The sub-leading terms arise from a variety of diagrams. It should be noted that the part of the above result proportional to  $k_{3,\mu} k_{3,\nu}$ is highly reminiscent of the analysis of the BCFW shift of two gluons in the bosonic string in \cite{Boels:2010bv}. There it was shown that this particular contribution is absent in the superstring. For the calculation here this can be checked as well: in the supersymmetric case this term is absent. 

\subsection{One scalar loop}
For the purposes of this article attention will now be further restricted to the terms which involve a scalar loop only. This is done since these terms generate so-called rational terms at the one loop level. With this motivation BCFW shifts of these contributions have been studied as such in the literature as recalled in the introduction. This corresponds for the leading term written above to setting $f^1_{0} =0$, as can be verified straightforwardly from equation \eqref{eq:loopa}. This leaves the analysis of the sub-leading terms, written in equations \eqref{eq:loopb} - \eqref{eq:looph}, suitably restricted to the scalar contributions. 

In a class of its own is the contribution of the remaining triangle integral, diagram b, given in \eqref{eq:loopb}. This is of order $z^0$ from the outset. Furthermore, it is easy to see that the results follow the same pattern as the term just analyzed. As such it is again proportional to the metric from the UV divergent part, proportional to $k_{3,\mu} k_{3,\nu}$ for the UV and IR finite part or antisymmetric in $\mu\nu$ from the IR divergent part. The UV and IR finite part arises from the scalar loop and entails a metric contraction between the soft legs. 

This leaves those integrals for which there is a non-trivial dependence on a shifted momentum in the loop integral. These diagrams can only contribute if they are UV divergent by the reasoning above. Note that this severely restricts the class of diagrams all the way down to four field terms. This is interesting in itself as it is known that the rational terms of the Yang-Mills amplitudes can be determined if the four field amplitude is known \cite{Bern:2005cq} \cite{Berger:2006ci}. 

Since one of the main interests in this section is to find deviations from the established pattern of \eqref{eq:largezform} one can disregard any contributions proportional to a metric contraction of the shifted legs. In addition, terms antisymmetric in the shifted legs can be dropped as well. Another way of enforcing this is to study shifts of adjacent \emph{like} helicity particles. As recalled in the appendix, the external wave functions for the $(++)$ and $(--)$ shifts scales as $\frac{1}{z}$ and $z$ respectively and are to leading order the same. 

One quickly discovers \emph{all} the terms of equations \eqref{eq:loopc} - \eqref{eq:looph} are proportional to the same pre-factor for the like helicity shifts,
\begin{align}
\epsilon^{\mu}_1 \epsilon_2^{\nu}  A^{(a)}_{\mu\nu\rho\sigma} \sim &  (D-2) \epsilon^{\mu}_1 \epsilon_2^{\nu}  \Gamma(\epsilon) \left(\frac{\Gamma(1-\epsilon)\Gamma(1-\epsilon)}{\Gamma(4-2\epsilon)} \right) \left(z \, q K_4\right)^{\epsilon} \eta_{\mu \sigma} \eta_{\nu \rho} [z^0] + \mathcal{O} \left(\frac{1}{z} \right) \\
& \equiv F_{\rho\sigma} \equiv F  \epsilon^{\mu}_1 \epsilon_2^{\nu}  \eta_{\mu \sigma} \eta_{\nu \rho} \ .
\end{align}
Here two overall factors $F$ and $F_{\rho \sigma}$ have been defined for notational convenience and $K_4$ is the momentum on one of the two soft legs. This result holds regardless of any on-shell conditions on the external legs other than the shifted ones. What remains is to add the six diagrams to determine the correct pre-factor. Since one can quickly check in the four point case that the above equation would give a non-zero contribution to a shift of the four point all helicity equal amplitude it is natural to suspect that, when summed, all diagrammatic contributions add up to a function proportional to $\epsilon$ to counter the UV pole. In fact, one arrives at the following list of results,
\begin{equation}
\begin{array}{ccccccc}
A^{c} & \rightarrow & -1 F \nonumber  & \qquad \qquad &
A^{d} & \rightarrow & \frac{1}{2} (3-2 \epsilon)(1-\epsilon) F \nonumber \\
A^{e} & \rightarrow & \frac{1}{4}(\epsilon - 1) F \nonumber  & \quad &
A^{f} &  \rightarrow & \frac{1}{8}(3-2\epsilon)(2-2\epsilon) F \nonumber \\
A^{g} & \rightarrow & -(3-2 \epsilon) F  \nonumber  & \quad &
A^{h} & \rightarrow & (2) F 
\end{array} 
\end{equation}
in $4-2\epsilon$ dimensions. When added, this yields
\begin{equation}\label{eq:whatsleft}
\epsilon^{\mu}_1 \epsilon_2^{\nu}  A^{(a)}_{\mu\nu\rho\sigma} = \frac{3}{2} F_{\rho\sigma} (\epsilon) (\epsilon - 1) \ , 
\end{equation}
confirming the earlier suspicion. A useful numerological trick to check consistency during the calculation is to track powers of $3$ appearing as coefficients for the $\left( \frac{1}{\epsilon} \right)$ poles: as these poles should cancel coefficients which do not multiply a three should cancel amongst themselves. This involves three instead of six diagrams. 

In summary, at leading $\left(z^1\right)$ and sub-leading order $\left(z^0\right)$ at one (scalar) loop there are only very few diagrams contributing as analyzed from the momentum integration. One class is given by gluing the triangle loop to a tree level gluon current, while the other consists of two tree level gluon currents glued to the four vertex calculated above. All these terms are manifestly finite. The leading order term arises from one diagram only.

\subsubsection*{Example: residues of helicity equal amplitudes}
The above can be used to compare to known results on BCFW shifts of single loop gluon amplitudes and in particular those in \cite{Bern:2005hs}. Here we will restrict to the helicity equal amplitudes. For definiteness and without loss of generality we pick $`+'$ helicity. The above results involve the tree level gluon current, which can be calculated in a variety of ways. For our purposes the all-plus gluon current as derived in \cite{berendsgiele} will suffice,
\begin{equation}
J_{+}^{\alpha\da}(3^+,4^+,\ldots, n^+)  \sim  \frac{q^{\da} q^{\db} (K)_{\db}^{\,\alpha}}{\braket{q 3} \braket{34}\ldots \braket{nq}} \ ,
\end{equation}
In this expression $q_{\da}$ is the dotted spinor associated to the light-like vector $q_{\mu}$ and $K$ is the momentum in the off-shell gluon. Combining this with the leading order triangle loop derived above, one immediately obtains 
\begin{equation}
\ldots \sim (k_1 - k_2 + z q)_{\rho} J_+^{\rho}= 0 \ ,
\end{equation}  
for any number of external legs. Hence for an adjacent shift of like helicity particles on the like helicity amplitude the scaling is of order $\left(z^0\right)$ and originates from the dressed four vertices discussed above. Also diagram (b) on the right hand side of figure \eqref{fig:1loopvertstriangles} whose value is calculated in equation \eqref{eq:loopb} does not contribute as a metric contraction between two soft all-plus currents vanishes. This leaves diagrams which involve equation \eqref{eq:whatsleft}. 

For definiteness, consider a BCFW shift for legs $1$ and $2$ for which the shift vector is
\begin{equation} 
q_{\a \da} = 1_{\a} 2_{\da} \ .
\end{equation}
Therefore the gauge reference spinor in the current has to be set to 
\begin{equation}
\xi^{\da} = 2^{\da} \ .
\end{equation}
The external wave functions of the shifted legs behave under the above shift as
\begin{align}
\frac{\xi_{\da} 1_{\a}}{\braket{\xi 1}} & \rightarrow \frac{\xi_{\da} 1_{\a}}{\braket{\xi (1 + 2 z}} \sim \frac{\xi_{\da} 1_{\a}}{z \braket{\xi  2}} + \mathcal{O}\left(\frac{1}{z^2}\right)  \\
\frac{\xi_{\da} 2_{\a}}{\braket{\xi 2}} & \rightarrow \frac{\xi_{\da} (2_{\a} - z 1_{\a})}{\braket{\xi 2}} \sim - z \frac{\xi_{\da} 1_{\a}}{\braket{\xi 2}} + \mathcal{O}\left(z^0\right) \ ,
\end{align}
for some arbitrary gauge spinor $\xi$ which may not be taken to be equal to $2_{\da}$. From the above the following form of the large $z$ shifts follows,
\begin{equation}\label{eq:allplusgenexpr}
A(z) \sim F z^0 \left(\sum_{j=3}^{n-1}  \frac{\xi_{\da} 1_{\a}}{\braket{\xi 2}} \frac{\xi_{\db} 1_{\b}}{\braket{\xi 2}} J^{\a \da}(3,\ldots,j) J^{\b \db}(j+1,\ldots,n)\right)  + \mathcal{O}\left(\frac{1}{z}\right)  \ . 
\end{equation}
It is easy to see that in this expression the $\xi$ dependence of the shifted wave functions drops out. In the case of four points there is only one term,
\begin{equation}
A(z) \sim F \left(\frac{\braket{13} \braket{14}}{\sbraket{23} \sbraket{24}}\right)  + \mathcal{O}\left(\frac{1}{z}\right) \ ,
\end{equation}
which is equivalent to
\begin{equation}
A(z) \sim - F \left(\frac{\braket{23} \braket{14}}{\sbraket{23} \sbraket{14}}\right)  + \mathcal{O}\left(\frac{1}{z}\right) \ .
\end{equation}
This is indeed the known residue at infinity of the adjacent shift of the four point all plus amplitude at one loop, up to and including all the dependence on $\epsilon$. This follow because the amplitude is proportional to the following integral
\begin{equation}
\int d^D(l) (\mu^2)^2 \frac{1}{l^2 (l+k_1)^2 (l+k_1+k_2)^2 (l-k_3)^2} \ ,
\end{equation}
where $\mu^2$ is the invariant length of the part of the dimensionally regulated loop momentum which is orthogonal to the four dimensions of choice. This is indeed proportional to $F$ in the limit, including all $\epsilon$ dependence up to the subleading term in the factor of $(D-2) = 2 - 2 \epsilon$. The latter discrepancy can be explained from $\epsilon$ scalar terms usually ignored in amplitude calculations at one loop.  In the four point case the residue at infinity coincides with the complete amplitude. 

For arbitrary multiplicity at leading order in $\epsilon$, the expression in equation \eqref{eq:allplusgenexpr} reads
\begin{equation}
A(z) \sim z^0 \left(\frac{1}{\braket{2 3} \braket{3 4} \ldots \braket{n 2}} \right)\left(\sum_{j=3}^{n-1} \sum_{k=3}^{j} \sum_{l=j+1}^{n} \frac{\sbraket{1k} \braket{k 2} \braket{j, j+1} \braket{2 l} \sbraket{l 2}}{\braket{2 j} \braket{j+1, 2}}\right)  + \mathcal{O}\left(\frac{1}{z}\right) +  \mathcal{O}\left(\epsilon\right) \ .
\end{equation}
Now for a fixed value of $k$, say $a$, and a fixed value of $l$, say $b$, there are only some specific terms which contribute to this expression
\begin{equation}
\left(\ldots k=a, l=b \ldots\right) =  \left(\frac{1}{\braket{2 3} \braket{3 4} \ldots \braket{n 2}} \right)\left(\sum_{j=a}^{b-1}  \frac{\sbraket{1a} \braket{a 2} \braket{j, j+1} \braket{2 b} \sbraket{b 2}}{\braket{2 j} \braket{j+1, 2}}\right) \ .
\end{equation}
This sum can be performed using 
\begin{equation}
\sum_{j=a}^{b-1} \frac{ \braket{j, j+1} }{\braket{2 j} \braket{j+1, 2}} = \frac{\braket{a b}}{\braket{2 a}\braket{b 2}} \ .
\end{equation}
Now the total contribution to the residue is the sum over all possible $a$ and $b$ such that $a<b$. This reads
\begin{equation} 
A(z) \sim z^0 \sum_{2<k<l\leq n} \left(\frac{\sbraket{1k} \braket{k l} \sbraket{l 1}}{\braket{2 3} \braket{3 4} \ldots \braket{n 2}  }\right)  + \mathcal{O}\left(\frac{1}{z}\right)  +  \mathcal{O}\left(\epsilon\right) \ .
\end{equation}
This can indeed be checked to be the residue at infinity as derived from the known expression at leading order in $\epsilon$. 

\subsection{Remarks}
It is quite plausible that the above derived large $z$ boundary terms are complete in the sense that for pure Yang-Mills these are the only boundary terms to be found. At leading order in $z$ this indeed follows from the background field analysis in this section. The sub-leading order $\left( z^0 \right)$ requires still more effort. However, all terms from the diagrams in figures \ref{fig:1loopvertstriangles}, \ref{fig:1loopvertswlightconepieces} and \ref{fig:1loopverts4pts}.
 which potentially contribute at order $\left(z^0\right)$ and which are not accounted for by the scalar contribution would contribute $\frac{1}{\epsilon}$ IR poles  to one of the purely rational amplitudes in pure Yang-Mills at large BCFW shift. Since it is known these are absent in the full answer they must cancel from other sources.  The divergences arising from Feynman parameter integrals have been left unevaluated until this point. As will be briefly indicated below, these divergences require care. 

For supersymmetric theories it is clear that the leading pole in $z$ at one loop at least has the same form as at tree level (being proportional to a metric). In fact, from the way this pole arises from the special three point vertex diagram it is expected that this can be proven fairly easily to all orders in perturbation theory. This follows since the form three particle coupling is very much restricted in Yang-Mills especially when supersymmetry is taken into account. Again, the sub-leading orders may be very complicated and the real problem is to show that there are no sources of $z$ dependence from all the other diagrams which interfere with this. For maximally supersymmetric theories knowledge of one good shift is enough to show that good super-shifts exist, see \cite{ArkaniHamed:2008gz}. 

Although a particular gauge setup was employed in deriving them there should be a more gauge invariant formulation of the same. After all, the full amplitudes are gauge invariant. For this one can note that additional BCFW shifts on other legs may be employed to derive auxiliary recursion relations for the order $z$ contributions. For the rational terms this is the strategy advocated in \cite{Bern:2005cq} \cite{Berger:2006ci}. What the above adds is a deeper understanding why these recursion relations work once the four point amplitude is known.

Note that the analysis gives a neat interpretation where the difference between cuts in four and higher dimensions is for pure Yang-Mills: if $\epsilon \rightarrow 0$ is taken first for cuts in four dimensions the rational parts are found at the residue at $z = \infty$. If the residue at $z = \infty$ is calculated first, the rational amplitudes are at the branch cuts. This quantifies the order of limits problem mentioned above. 

\subsubsection*{Improved shifts arising from IR divergences}
From the massless box example one can see how terms which have been counted as sub-leading in $z$ up until now may be elevated. The point is that if possible divergences are encountered when deriving the Mellin-Barnes representation, there will be an enhancement of the $z$ scaling. See for instance equation \eqref{eq:telltalefeautreoflift} and the discussion below which illustrates this point neatly. On the other hand, for the massless box type integral this only arises if the exponent $i$ in 
\begin{equation}
\int da_4 \frac{1}{(\alpha_1 \alpha_3 s + \alpha_2 \alpha_4 t)^i} \ , 
\end{equation}
is bigger than $2$ and there are no redeeming powers of $\alpha_i$ around to dampen the divergence. Moreover from inspecting the poles passing the contour this effect can only elevate a naively order $z^{-i}$ suppressed integral to $\frac{1}{z}$. Hence this mechanism only contributes at higher order in $z$ if there is another, explicit power of $z$ floating around in the integral which can be used to generate a contribution and the integral is sufficiently IR infinite. This cannot happen for the integrals appearing in the scalar loop. For the complete background field calculation for the gluon in the loop note that for the leading term in $z$ this can at most contribute a metric as the two possible sources of $z$ in the numerator of the integrand are the momenta at legs $1$ and $2$. To yield a $z = (\frac{1}{z}) \times z^2$ contribution these have to be contracted somewhere else in the diagram and not to each other. This leaves the metric contraction between the shifted legs. Intriguingly, this term is proportional to $\frac{1}{\epsilon^2}$ whose numerator on general grounds is expected to be related to the tree amplitude again and it is encouraging that the leading poles indeed conform to this expectation. 

A full discussion of IR divergences will be deferred to future work. For inspiration see \cite{Friot:2005cu}  and references therein. 

\section{Towards BCFW on-shell recursion relations at loop level}
\label{sec:reltoamp}
Although this article mainly deals with the large BCFW shift behavior of the integrand and its integrals one of the prime objectives of this line of research is to obtain similarly useful on-shell recursion formulae at loop level as exist at tree level. In this section some (very) preliminary developments in this direction are presented. The main question is what the singularities of the integrand and the integrals correspond to in terms of lower loop or lower point amplitudes. Since branch cut singularities in the integral generically correspond to a cut propagator in the integrand these questions are essentially the same, up to the discrepancy quantified in the previous section. This will be ignored here as it is mostly relevant when discussing the dimension in which the cuts are calculated. 

From the good large $z$ behavior for color-adjacent shifts one can write a prototype recursion relation for any amplitude. 
\begin{equation}\label{eq:recursionslooplevelgenguess}
A^{(l)}_n(1,2,3 \ldots, n) = A_{\textrm{tree poles}}^{(l)} + A^{(l-1)}_{cut} \ ,
\end{equation}
where the superscript $l$ in $A^{(l)}$ denotes the loop order. Here particles one and two have been shifted, which are assumed to be color adjacent. The first term arises from poles in tree parts of the amplitudes and are just a natural generalization of tree level BCFW, i.e.
\begin{equation}
A_{\textrm{tree poles}}^{(l)} = \sum_{r,h(r)} \sum_{i=0}^l \sum_{k=3}^{n-1}
\frac{A^{(l-i)}_{k}(\hat{2}, \ldots, k, \hat{P}_r)
A^{(i)}_{n-k+2}(\hat{P}_r, k+1, \ldots, \hat{1})}{\left(k_2 +
\ldots k_k \right)^2 } \ ,
\end{equation}
where the first sum is over all different mass levels $r$ and over all polarization states at that level, denoted $h(r)$. The new term arises from single cuts of the amplitude and reads
\begin{equation}
A^{(l-1)}_{cut}  = \sum_{c,r, h(r)} \int d^{D} l\,\, \left( \frac{1}{l^2} \right) B^{(l-1)}\left(\hat{1}, \hat{l}, -\hat{l}, \hat{2},3 \ldots, n \right)  \ ,
\end{equation}
where $c$ sums over the possible single cuts and $\hat{l}$ is on-shell,
\begin{equation}
\hat{l} = l - q \frac{l^2}{2 l \cdot q} \ .
\end{equation}
The question therefore is if there is an interpretation of the quantity $B^{(l-1)}$ in terms of lower loop, higher point amplitudes. A natural guess would be
\begin{equation}
B^{(l-1)}_{n+2}\left(\hat{1}, \hat{l}, -\hat{l}, \hat{2},3 \ldots, n \right) \qquad  \stackrel{\Large{\bf ?}}{\longleftrightarrow} \qquad  A^{(l-1)}_{n+2}\left(\hat{1}, \hat{l}, -\hat{l}, \hat{2},3 \ldots, n \right) \ .
\end{equation}
Note however that from the diagrams it is easy to see that $B$ is finite while the term on the right hand side would be evaluated at a collinear singularity. More cunning is therefore required to resolve this issue. The simplest suggestion would be to simply subtract of the collinear pole from the lower loop amplitude in some regularized fashion as it is basically the product of two amplitudes again, but this would leave a soft pole to deal with. See \cite{CaronHuot:2010zt} for some more suggestions how to define this `forward limit' properly in a very much  related context. Any of these suggestions immediately leads to a concrete recursion relation. Note that the authors of \cite{NigelGlover:2008ur} sidestep the issue by working only with those MHV diagrams which do not have tadpoles. This is another way of defining the quantity $B$ to any loop order (see also \cite{Sever:2009aa}) and would express the tree amplitude in terms of tree amplitudes again.

\FIGURE[!h]{\epsfig{file=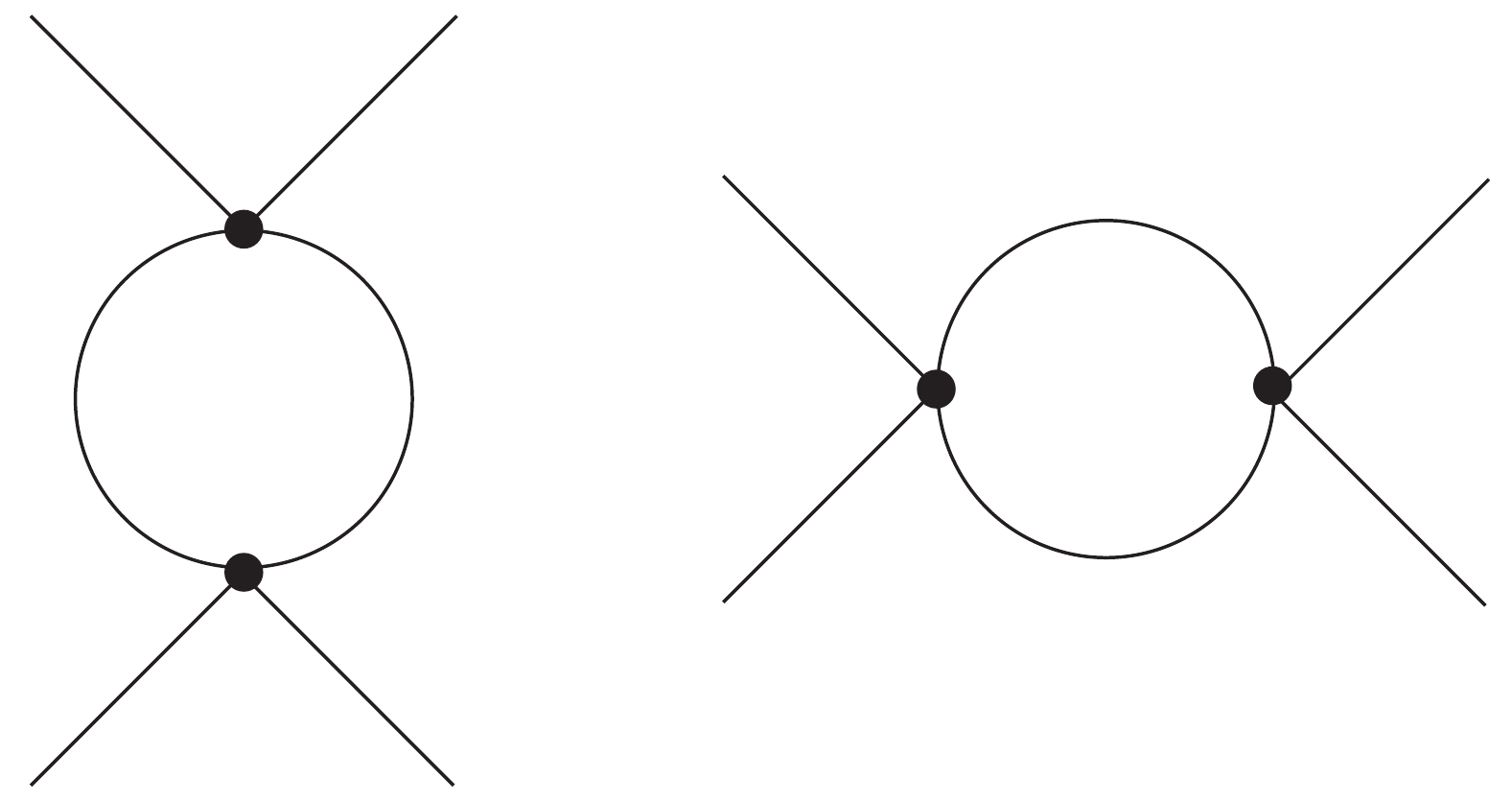,scale=0.55} \caption{Diagram topologies contributing to the four point scattering amplitude at one loop in $\phi^4$ theory} \label{fig:phi4oneloop}}

Some further inspiration can be taken from  $\phi^4$ theory. There the four particle scattering amplitude for instance has two diagrammatic contributions at one loop as depicted in figure \ref{fig:phi4oneloop}. Sums over different particle orderings will be suppressed. BCFW shifting particles on two adjacent legs gives a $\left(z^0\right)$ behavior of the integrand, exactly as at tree level. This can be sidestepped by introducing more particles which modify the UV behavior of the theory \cite{Boels:2010mj} but what is important here is the pole of the integrand which does get a shift. Cutting the associated line gives a tree structure, obviously. However, this is not a $6$ particle tree level amplitude: what is missing for this is the Feynman graph which has the cut legs attaching to the same vertex on the tree. This tree diagram can be added and subtracted schematically as,
\begin{equation}
(A_1^4)\left(\{p_i\}\right) = \int \frac{d^D l}{(2 \pi)^D} \frac{\delta^D(l^2 + z q k_1 - m^2)}{l^2- m^2} A_0^6\left(\{\tilde{p}_i\}, \tilde{l}, -\tilde{l}\right) - F\left(\{\tilde{p}_i\}, \tilde{l}, -\tilde{l}\right) \ .
\end{equation}
When employed within the recursion approach the integral over the second term would vanish in dimensional regularization as it is a tadpole type integral. For Yang-Mills this suggests to sum only those diagrams which do not have the cut legs attached to the same vertex, which naively coincides with the prescription in \cite{NigelGlover:2008ur}. 

\subsection*{Nested shifts}
One obvious suggestion to sidestep the forward limit issue is to use an auxiliary BCFW shift on two adjacent legs on the residue of the single cut. If the residue is gauge invariant up to terms which integrate to zero, one can use the same argument as above to argue that the single cut residue  again scales as a shift of a tree level Yang-Mills amplitude to all loop orders. Now at one loop for instance this reduces the single cut residue to either a more familiar double-cut residue or contributions which involve $B$ type terms with less legs dressed by tree amplitudes. At higher loops part of the terms obtained will be double-cuts while others will be two single cuts of separate loops. Iterating shifts should eventually yield an expression of the amplitude in terms of integrals over tree level amplitudes evaluated at shifted momenta. 

Schematically this reads at one loop,
\begin{equation}
B^{(0)}_{n+2}\left(\hat{1}, \hat{l}, -\hat{l}, \hat{2},3 \ldots, n \right) =   \sum_{r,h(r)} \sum_{k=3}^{n-1}
\frac{C^{0}_{k}(\ldots, \widetilde{n-1}, \widetilde{P}_r)
C^{0}_{n-k+2}(\widetilde{P}_r, \widetilde{n}, \ldots)}{\left(K \right)^2 } \ ,
\end{equation}
where the tildes denote the second BCFW shift, taken on particles $n$ and $n-1$. The quantity $C$ is either an $A$ or a $B$ depending on the location of the $\tilde{l}$ legs: if both of them are on a certain $C$ it is a $B$-type, if one of them is both $C's$ are an $A$ and if none of them are on a certain $C$ it is an $A$-type. Now equation \eqref{eq:recursionslooplevelgenguess} can be used in reverse to express the $B$-type terms on the right hand side again in terms of lower (maximally $n-1$) point one loop amplitudes. This expresses a one loop scattering amplitude in terms of scattering amplitudes only with either a lower loop or a lower leg order. 

The resulting expressions are rather messy and at higher loops the involved structures for nested cuts get more and more intricate. It will be very interesting to see if more efficient formulas than this can be made to work, especially when combined with maximal supersymmetry.  Note that in deriving the formulas above several issues with order of limits problems between the two shifts and a potential problem with assumed gauge invariance have been neglected: these most certainly deserve further study. 

\section{Discussion and conclusions}
In this article a first step was made towards on-shell recursion at loop level. It was proven the Yang-Mills integrand can be reconstructed from its singularities in full analogy with tree amplitudes by studying large BCFW shifts of the integrands. Clearly, the shifts are probing a form of universal behavior in Yang-Mills theory. The main obstacle to apply on-shell recursion relations is to find a proper interpretation of the singularities of the integrand in terms of integrands for lower loop or lower point amplitudes. The preliminary analysis above for double shifts leaves much to be desired and clearly much more interesting work can and needs to be done here.

It should also be very interesting to extend the analysis of the difference between integrand and integral made explicit above at one loop. Especially higher loops would be most welcome. That this should be possible follows from the link to UV and IR singularities of the integrand. For UV singularities at one loop a full discussion was given above.  The next step here is to understand the shift behavior of the Feynman parameter integrals at one loop in more technical detail, which should enable the natural extension of the analysis to two loops and beyond. It is clear that the number of Feynman graphs to be calculated is going to be limited, especially compared to usual two-loop computations. This should yield already some interesting new results for the general form of two-loop amplitudes. Note that the analysis of large BCFW shifts gives information on complete amplitudes which is in a real sense complimentary to the usual unitarity cuts. It would also be interesting to see if this information can already be used to explicitly calculate amplitudes. For instance, the large z behavior obtained above could be used in principle to constrain the rational terms to higher (all) orders in $\epsilon$. 

From the results above it is easy to speculate that the Einstein gravity amplitude integrand will obey the same large-$z$ scaling behavior as at the tree level (i.e. the `square' of that of Yang-Mills theory). This is certainly plausible both from the string theory point of view as well as tree and loop level experience. This leads one to suspect for instance that loop level relations for integrands like those conjectured in \cite{Bern:2010ue} between gauge theory and gravity can be proven just as the tree level version of these relations \cite{Bern:2008qj} can be proven from non-adjacent BCFW shifts \cite{Feng:2010my} (see also \cite{BjerrumBohr:2010yc}). In this respect it is encouraging that non-adjacent shifts of particles on the same color trace for Yang-Mills amplitudes are suppressed by $\left(\frac{1}{z}\right)$ for all cases considered in this article by simple absence of the leading pole diagrams. This implies the existence of loop level `bonus' relations. Moreover, the results on shifts obtained in this article is independent of the color structure of the unshifted parts of the diagrams: they hold in particular for non-planar amplitudes. 

It will be interesting to study the application of the ideas of on-shell recursion at loop level to maximally supersymmetric Yang-Mills theory in four dimensions as it opens a window to prove conjectured properties of the integrals and integrands such as dual conformal invariance and its one-loop breaking (see \cite{Henn:2009bd} and references therein and thereto). Also, there should be a close link to on-shell recursion in twistor spaces and the conjectured Grassmanian structure of the leading singularities of the amplitudes in maximally supersymmetric Yang-Mills. It is exceptionally natural to transform the above recursion formulae and its natural generalizations to super twistor and super ambi-twistor space in the light of the results in \cite{Mason:2009sa} and \cite{ArkaniHamed:2009si}. This is in a real sense the same intuition which motivated the present paper (see subsection \ref{subsec:stringmot}): on-shell recursion should be related to a form of the CFT bootstrap equations also at loop level. Twistor spaces are after all closely connected to the conformal nature of $\mathcal{N}=4$ super-Yang-Mills. Actually this is not too far off the idea that also at weak coupling maximally supersymmetric Yang-Mills should be related to a string theory \cite{witten}. This can perhaps be made more precise starting from the tree level results in \cite{Skinner:2010cz}.

Finally, it should be noted that there no inherent obstructions to the loop order of the amplitudes under study for the techniques developed here. This opens the window to NNLO calculations in QCD for instance an admittedly very small fraction. Certainly first the one-loop structure needs to be under full control and cross-checked to known results before any serious attack can be made on experimentally relevant calculations beyond the one loop level. However, as methods beyond one loop for standard model calculations are currently limited to delicately applied brute force even a slightly opened window is a tempting prospect. 

\section*{Acknowledgements}
It is a pleasure to thank David Skinner and James Drummond for discussions, as well as the anonymous referee for several suggestions to improve the exposition. Figures in this article have been made using Jaxodraw \cite{Binosi:2008ig}. This work was supported by the German Science Foundation (DFG) within the Collaborative Research Center 676 "Particles, Strings and the Early Universe". 

\appendix
\section{BCFW shifts of fermions in $D$ dimensions}
\label{app:BCFWconv} 
In this appendix the analysis of shifts of gluons for the Yang-Mills integrand in the main text is extended to shifts of external fermions in $D$ dimensions. In addition this appendix spells out some conventions about quantum numbers and the BCFW shift. The starting point is
\begin{align}
k_1 \rightarrow & \,\,\, k_1 + q z \\
k_{2} \rightarrow  & \,\,\, k_2 - q z \ ,
\end{align}
for two particles labelled $1$ and $2$. As noted in the main text, the following vectors span a four dimensional subspace,
\begin{equation}\label{eq:specfourframe}
\textrm{Span} \left(q, \bar{q}, k_1, k_2 \right) \ ,
\end{equation}
of a $D$-dimensional space. In other words, there is a choice of normalization of $q$ and a choice of Lorentz frame such that
\begin{align}\label{eq:specframe}
q & = \frac{1}{\sqrt{2}} (0, 1, i, 0, \ldots)\\
\bar{q} & = \frac{1}{\sqrt{2}} (0, 1, -i, 0, \ldots) \\
k_1 & = \frac{k}{\sqrt{2}} (1,0,\ldots ,1)\\
k_2 & = \frac{k}{\sqrt{2}} (1,0,\ldots,-1)  \ .
\end{align}
Moreover, a choice of ordering of the entries has been used to make the formulae more readable. These equations are for instance the starting point of the higher dimensional analysis of BCFW shifts in \cite{ArkaniHamed:2008yf}.  From the explicit expressions it is easy to see that $q$ and $\bar{q}$ are helicity eigenstates of momenta $k_1$ and $k_2$, where helicity is defined w.r.t. the four dimensional subspace, i.e.
\begin{equation}\label{eq:fourdimgen}
R_1 = \frac{k^{\mu}_1 k^{\nu}_2 \epsilon^{(4)}_{\mu\nu\rho\sigma} \Sigma^{\rho \sigma}}{2 k_1 \cdot k_2} \ ,
\end{equation}
where $\epsilon^{(4)}$ is the totally antisymmetric symbol in the subspace spanned by the vectors in \eqref{eq:specfourframe} and $\Sigma$ is the generator of rotations in $D$ dimensions. Note that $\epsilon^{(4)}$ depends non-trivially on the momenta in general. Also, this is the helicity operator for $k_2$ while it is \emph{minus} the helicity operator for $k_1$. 

It is very natural to take the momentum of the other leg as the choice of gauge for each of the legs. The polarizations of the gluon are then given by $q$ and $\bar{q}$ and can be split up into either nontrivial ($\pm$) or trivial (transverse) under the above helicity generator. Under the BCFW shift these obey
\begin{equation}
\label{eq:shift_rules}
\begin{array}{ccccccc}
g_1^- & = & g_2^+ & = & q   & \rightarrow & q\\
        &   & g_1^+ & = & \bar{q} & \rightarrow & q^*+ \frac{z}{k} \, k_2 \\
        &   & g_2^- & = & \bar{q} & \rightarrow & q^*- \frac{z}{k} \, k_1 \\
        &   &         &   & g^T & \rightarrow &g^T
\end{array}  \ .
\end{equation}
This concludes the analysis of the external wave functions for gluons. What is needed below in addition to this are the polarization spinors. These are solutions to the Dirac equation for either $k_1$ or $k_2$ with definite eigenvalues under the generator \eqref{eq:fourdimgen}. In the special four dimensional frame these can be constructed using the four dimensional spinor helicity method. However, in higher dimensions in general it will pay to be slightly more general, this will be done below.

The main observation needed to study the spinor polarizations is that one can construct the set
\begin{equation}\label{eq:complbasislight}
\left\{ q, \bar{q}, k_1, k_2, g^T_i \bar{g}^T_i \right\} \ ,
\end{equation} 
with all vectors light-like to form a basis of $\mathbb{R}^D$ with $D$ even. Here the transverse polarization vectors have been grouped into (normalized) conjugate pairs,
\begin{equation}
g^T_i \cdot \bar{g}^T_j = \delta_{ij}\quad \qquad i,j \in \{1,\ldots, D-4 \} \ .
\end{equation}
In passing we note the relation of the set just constructed to the choice of complex structure on $\mathbb{R}^D$ which can be seen most easily from equation \eqref{eq:specframe}. Using the basis of \eqref{eq:complbasislight} one can study representations of the $\gamma$ matrix algebra,
\begin{equation}
\{\gamma_{\mu}, \gamma_{\nu} \} = 2 \eta_{\mu\nu} \ ,
\end{equation}
by projecting. Define for this the operators
\begin{equation}
\begin{array}{ccccc}
\Gamma^+_0 & = \frac{1}{k \sqrt{2}}  k_1^{\mu} \gamma_{\mu} & \qquad & \Gamma^-_0 & = \frac{1}{k \sqrt{2}}  k_2^{\mu} \gamma_{\mu}  \\
\Gamma^+_1 & = \frac{1}{\sqrt{2}}  q^{\mu} \gamma_{\mu} & \qquad &  \Gamma^-_1 & = \frac{1}{\sqrt{2}}  \bar{q}^{\mu} \gamma_{\mu}\\
\Gamma^+_{i+1} & = \frac{1}{\sqrt{2}}  g^T_i \gamma_{\mu} & \qquad & \Gamma^-_{i+1} & = \frac{1}{\sqrt{2}} \bar{g}^T_i \gamma_{\mu}
\end{array} \ ,
\end{equation}
so that the Clifford algebra reduces to $D/2$ copies of the (normalized) fermionic harmonic oscillator,
\begin{equation}
\{\Gamma^+_i, \Gamma^-_j\} = 1 \ .
\end{equation}
The standard representation theory of this algebra can be written neatly in terms of the half-integer eigenvalues under the `rotation' generator
\begin{equation}
R_i = \Gamma^-_i \Gamma^+_i - \frac{1}{2} \ .
\end{equation}
Let the vector of eigenvalues be $\vec{h}$. All spinors are annihilated by half of the generators, while the other half of the generators flips one eigenvalue,
\begin{equation}
\Gamma_i^{\pm} | \vec{h} \rangle = \delta_{h_i, \mp \frac{1}{2}} |\vec{h}' \rangle \ ,
\end{equation}
where $\vec{h}'$ has the $i$-th eigenvalue (labelled $h_i$) inverted. The chirality of the spinor is the product of the signs of the eigenvalue labels. Of course, the first eigenvalue label on the spinors corresponds to the spinor being a solution to the Dirac equation for $k_1$ for $h_0 = +\frac{1}{2}$ or $k_2$ for $h_0 = -\frac{1}{2}$. The constructed spinors form a basis of the space of all spinors. 

With this construction in hand, it is now straightforward to study the BCFW shift considered in this article. The shift concerns only the first two quantum numbers of the spinors as expected and leads to the following shifted spinors
\begin{equation}
\begin{array}{ccc}
|\frac{1}{2}, \frac{1}{2}, \ldots \rangle & \rightarrow & |\frac{1}{2}, \frac{1}{2}, \ldots \rangle  \\
|\frac{1}{2}, - \frac{1}{2}, \ldots \rangle & \rightarrow &  |\frac{1}{2}, - \frac{1}{2}, \ldots \rangle - \frac{z}{k} \,\,|-\frac{1}{2},  \frac{1}{2}, \ldots \rangle \\
|-\frac{1}{2}, - \frac{1}{2}, \ldots \rangle & \rightarrow & |-\frac{1}{2}, - \frac{1}{2}, \ldots \rangle + \frac{z}{k} \,\,|\frac{1}{2},  \frac{1}{2}, \ldots \rangle \\
 |-\frac{1}{2},  \frac{1}{2}, \ldots \rangle & \rightarrow &  |-\frac{1}{2},  \frac{1}{2}, \ldots \rangle  
\end{array} \ .
\end{equation}

It is not too hard to verify that expressed in terms of four dimensional spinor helicity this is the usual BCFW shift. The four dimensions of course correspond to the special four dimensional space spanned by $k_1,k_2,q,\bar{q}$ and the first two quantum number labels on the spinors. The four dimensional chirality/helicity of the spinors for instance is the product of the signs of these two labels. The remaining labels can be interpreted as the R-symmetry labels in this four dimensional decomposition. The four dimensional interpretation can hence be captured in the following notation
\begin{equation}
\begin{array}{ccc}
\psi^-_{I} (k_1) & \rightarrow & \psi^-_{I}(k_1)   \\
\psi^+_{I} (k_1)   & \rightarrow & \psi^+_{I}(k_1)  - \frac{z}{k} \,\,\psi^+_{I}(k_2)  \\
\psi^-_{I} (k_2) & \rightarrow & \psi^-_{I}(k_2)  + \frac{z}{k} \,\,\psi^-_{I}(k_1)  \\
\psi^+_{I} (k_2)  & \rightarrow &  \psi^+_{I}(k_2)   
\end{array} \ ,
\end{equation}
where the R-symmetry indices should be interpreted as a $(D-4)/2$ dimensional vector with half-integer entries. This is the notation used below. 

For completeness the shifts of the conjugate spinors can be derived from the completeness relation,
\begin{equation}
\gamma_{\mu} (k_1 + z q)^{\mu} = \sum_{h=\pm} \tilde{\psi}_{\pm}(k_1) \overline{\tilde{\psi}_{\pm}(k_1)} \ .
\end{equation}
In particular,
\begin{equation}
\begin{array}{ccc}
\overline{\psi^+_{I} (k_1)} & \rightarrow & \overline{\psi^+_{I}(k_1)}   \\
\overline{\psi^-_{I} (k_1)}   & \rightarrow & \overline{\psi^-_{I}(k_1)}  - \frac{z}{k} \,\,\overline{\psi^-_{I}(k_2)}  \\
\overline{\psi^+_{I} (k_2)} & \rightarrow & \overline{\psi^+_{I}(k_2)}  + \overline{\frac{z}{k} \,\,\psi^+_{I}(k_1)}  \\
\overline{\psi^-_{I} (k_2)}  & \rightarrow &  \overline{\psi^-_{I}(k_2)}   
\end{array}
\end{equation}
is obtained through this. The conjugate spinors are the solution to the conjugate Dirac equations. The above analysis is an application of techniques explored in \cite{Boels:2009bv}. Indeed, the above can easily be turned into a completely covariant treatment by replacing
\begin{equation}
k \rightarrow \sqrt{k_1 \cdot k_2} 
\end{equation}
everywhere and phrasing the basis in terms of its inner products only. 

\subsection*{BCFW shifts involving one or two fermionic legs}
Now the stage is set to extend the reasoning in the main text to shifts of one gluon and one fermion and two fermion legs in generic minimally coupled (renormalizable) gauge theories. For shifts involving a fermion a choice of gauge is not necessary as the na\"ive powercounting can be used to derive the large $z$ behavior, as shown below. For fermions the analysis below makes the argument touched upon in \cite{Cheung:2008dn} explicit. 

\subsubsection*{Two shifted gluons}
As shown in the main text for two gluons the large $z$ behavior of the amplitude follows from
\boxit{\begin{equation}\label{eq:largezansatz}\
A = \epsilon_{1}^{\mu}\left(\hat{k}_1\right) \epsilon_{2}^{\nu}\left(\hat{k}_2\right) M_{\mu\nu}  = 
\epsilon_{1}^{\mu}\left(z \right) \epsilon_{2}^{\nu}\left(z\right) \left(f_1 \eta_{\mu\nu} z + B_{\mu\nu} + \mathcal{O}\left(\frac{1}{z}\right) \right) \ .
\end{equation}}
Here $\epsilon_1$ and $\epsilon_2$ are the polarization vectors of the shifted adjacent legs, the hats on the momenta denote shifted momenta and $f_1$ is an arbitrary function of the unshifted momenta and polarization vectors. The matrix $B$ is antisymmetric. This Ansatz follows from the above argument about the class of dominating diagrams. Combining the large $z$-behavior of the polarization vectors with the Ansatz yields table \ref{tab:largezn4}. 

\subsubsection*{One shifted gluon, one shifted fermion}
In general diagrams containing external fermion legs have a definite ordering of the fermion lines which has to be picked. This determines the external fermion wave functions to be given by either $\psi$ or its conjugate $\overline{\psi}$. The quantum numbers of the fermions can be selected too, see the appendix for a construction. 

In this case the large $z$ behavior of the amplitude follows from
\boxit{\begin{multline}\label{eq:onefermonegluon}
A = \epsilon_{1}^{\mu}\left(\hat{k}_1\right) M_{\mu} u\left(\hat{k}_2\right)  = 
\epsilon_{1}^{\sigma}\left(z\right) \left(\eta_{\sigma}^{\mu}  + K_{\sigma}q^{\mu} + q_{\sigma} \tilde{K}^{\mu} + f_1 q_{\mu} q^{\sigma} \right)\\    \left(\overline{\psi} \left[\gamma_{\mu} (1+ \hat{K}^{\rho} \gamma_{\rho} q_{\nu} \gamma^{\nu} \right]+ \mathcal{O}\left(\frac{1}{z}\right) \right) u\left(z\right) \ ,
\end{multline}}
where $u(k_2)$ is the (in general $z$-dependent) polarization spinor of the fermion, $\overline{\psi}$ is an arbitrary (conjugate) spinor, $f_1$ is an arbitrary function and $K, \tilde{K}$ and $\hat{K}$ are arbitrary vectors. 

In general the hard line connecting the external particles contain scalar fields, multiple fermion fields and gluonic contributions. The leading contribution is again a combination of very simple three point vertices. Combinations of hard gluon and scalar lines coupling to this glue have been analyzed above. Note that possible vertices with the Yukawa couplings between scalars and fermions are sub-leading in the large z limit as they are momentum independent. There is at least one fermion attaching to the hard line since this connects to the external particle. This fermion vertex lowers the overall $z$ count by one. The vertex to which the shifted gluon leg attaches directly on the hard line contains one gamma matrix in the complete gamma matrix trace, whose structure can be denoted as
\begin{equation}
 \epsilon_{1}^{\nu}  \left(\overline{\psi}  \gamma_{\nu} \left(q \!\!\! \slash \right) \gamma_{\mu_2} \ldots \left(q \!\!\! \slash \right) \right) u(k_2) \ .
\end{equation}
In this expression the $\mu_i$ indices on the gamma matrices attach to other parts of the diagram which are left undetermined as they will not influence the final result. Since $\left(q \!\!\! \slash\right)^2 =0$ this reduces to 
\begin{equation}
\left(\overline{\psi}  \gamma_{\nu}  \left(1 + \hat{K}\!\!\! \slash q \!\!\! \slash \right)   \right) u(k_2) \left(\prod_i q_{\mu_i} \right) \ ,
\end{equation}
by repeated use of the gamma matrix algebra. Additional fermions along the gluonic hard line will lead to an additional suppression of $\frac{1}{z}$ for diagrammatic reasons. Assembling the hard gluon line contribution with the above coupling to the fermionic part of the hard line now leads to the equation \eqref{eq:onefermonegluon}. The large $z$-behavior of the amplitudes which follows from the ansatz will be listed below in table \ref{tab:largezferms}. 

\subsubsection*{Two shifted fermions}
For the case of two shifted fermions the large $z$ behavior of the amplitude follows from the following $2$ particle current, 
\boxit{\begin{equation}
A = \overline{u\left(\hat{k}_1\right)} M u\left(\hat{k}_2\right) = \overline{u\left(z\right)} \left(K^1_{\mu} \gamma^{\mu} + \frac{1}{z} K^2 + \mathcal{O}\left(\frac{1}{z^2}\right) \right) u\left(z\right) \ ,
\end{equation}}
which follows directly from power-counting the involved diagrams. The $z$-independent but otherwise arbitrary functions $K^1_{\mu}$ and $K^2$ can only contain an even number of gamma matrices. The first arises from a fermionic hard line while the second contains one gluon exchange diagram. 

The large $z$ behavior which follows from the above analysis of shifted fermions is captured in table \ref{tab:largezferms}. The entries containing one fermion and one gluon in this table were discussed in \cite{Cheung:2008dn}. In the table the fermion polarizations are listed as defined in a special set of $4$ dimensions and split into helicity and R-symmetry labels as explained more fully above. The difference between the entries for $g^T$ and $g^{T2}$ on one axis and fermions on the other is that
\begin{equation}
g^{T}_{\mu} \gamma^{\mu} \psi^{\pm} \neq 0 \quad \textrm{while} \quad g^{T2}_{\mu} \gamma^{\mu} \psi^{\pm} = 0 \ ,
\end{equation}
while the two different entries for the fermion shifts correspond to either the same R-symmetry labels ($1$) or different ones ($0$). Note that the fermion entries in this table were derived by straightforward powercounting, while only the sub-leading behavior of the gluon shifts required more care in the form of the above AHK gauge powercounting. 

\begin{table}
\begin{center}
\begin{tabular}{c|c c c c c}
$\epsilon_1 \;\backslash \;\epsilon_2  $ & $g^-$              & $g^+$              & $g^T$  & $\psi^-_{I}$ & $\psi^+_{I}$ \\
\hline
$g^-$                   & $ +1$ & $ +1$ & $ +1$  & $ +1 $ & $ +1 $  \\
$g^+$                   & $ -3$ & $ +1$ & $ -1$ & $ -2 $ & $ 0 $  \\
$g^T$                     & $ -1$ & $ +1$ & $ -1$ & $ -1 $ & $ 0 $  \\
$g^{T2}$                    & $ -1$ & $ +1$ & $0$ & $ 0 $ & $ +1 $  \\
$\psi^-_{J}$          & $ 0 $ & $ +1 $ & $0/-1$ & $ +1/0 $ & $ 0 $  \\
$\psi^+_{J}$          & $ -2$ & $ +1$ & $0/-1$ & $ -2 $ & $ +1/0 $ 
\end{tabular}

\caption{leading asymptotic power in $z^{-\kappa}$ of the adjacent BCFW shift of two particles in a tree amplitude in $D$ dimensions for all possible polarizations}
 \label{tab:largezferms}
\end{center}
\end{table}

As a consistency check  at tree level one can use the supersymmetric Ward identities to check that if the last two columns of table \ref{tab:largezferms} are known, the other entries may be derived. The power of the supersymmetric Ward identity is that it is independent of the coupling constants in the theory, so these relations must hold for amplitudes at any loop level in Yang-Mills theory.

\section{One loop diagrams contributing to the BCFW shift at leading order}
\label{app:loopintegrals}
In this appendix the explicit results for the one loop diagram contributions as obtained through the color ordered background field method used in the setup in the main text are collected. These have been calculated with the color ordered Feynman rules as can be found in figure (8) in \cite{Bern:1996je}. Legs one and two will be taken to be shifted. Sub-leading terms in the BCFW shift parameter $z$ will be discarded as indicated. Subscripts on square brackets indicate how many powers of the background field  type vertex coupling two quantum fields to the background field field strength have been used. Schematically this reads
\begin{equation}
\left( a_{\mu} F^{\mu\nu}[A] a_{\nu}\right)^{\bf i}  \qquad \leftrightarrow \qquad \left[\ldots \right]_{\bf i} \ .
\end{equation}
The diagrams themselves are listed in figures \ref{fig:1loopvertstriangles}, \ref{fig:1loopvertswlightconepieces} and \ref{fig:1loopverts4pts}.

\subsection*{Triangle vertices with two shifted legs}
This class of diagrams is illustrated in figure \ref{fig:1loopvertstriangles}. The trivalent triangle vertex evaluates to
\begin{multline}\label{eq:loopa}
\epsilon^{\mu}_1 \epsilon_2^{\nu}  A^{(a)}_{\mu\nu\rho} = \sqrt{2} \epsilon^{\mu}_1 \epsilon_2^{\nu}  \int \frac{d^D l}{(2 \pi)^D}  \,\, \frac{1}{l^2(l-k_1)^2 (l+k_2)^2} \left((D-2) \left[l_{\mu} l_{\nu} (2l + k_2 - k_1)_{\rho} \right]_0  \right. \\ \left. + 4\left[(l_{\mu} \eta_{\rho\nu} + l_{\nu} \eta_{\rho\mu} - l_{\rho} \eta_{\mu\nu})(k_1 \cdot k_2) \right]_{2/3} \right) + \mathcal{O} \left(\frac{1}{z} \right) \ .
 \end{multline}

\noindent The triangle vertex with two shifted legs and two external legs evaluates to
\begin{multline}\label{eq:loopb}
\epsilon^{\mu}_1 \epsilon_2^{\nu}  A^{(b)}_{\mu\nu\rho} = - \epsilon^{\mu}_1 \epsilon_2^{\nu} \int \frac{d^D l}{(2 \pi)^D} \frac{1}{l^2(l-k_1)^2 (l+k_2)^2}\left[(D-2) l_{\mu} l_{\nu} \eta_{\rho\sigma}\right]_0 + \\
2 \left[k_{\mu} k_{\nu} - (k_1 \cdot k_2) \eta_{\mu\nu} + 4 l_\mu \left(k_{2,\rho} \eta_{\nu\sigma} -k_{2,\sigma} \eta_{\nu\rho}\right) + 4  l_\nu \left(k_{1,\rho} \eta_{\mu\sigma} -k_{1,\sigma} \eta_{\mu\rho}\right)  \right]_2 \\
+ 2 \left[ (k_1 \cdot k_2) (\eta_{\mu\rho} \eta_{\nu\sigma} -\eta_{\nu\rho} \eta_{\mu\sigma}) + \eta_{\mu\nu} (k_{2,\rho} k_{1,\sigma} - k_{2,\sigma} k_{1,\rho}  ) + \right.\\ \left.k_{\nu} (\eta_{\sigma \mu} k_{2,\rho} - \eta_{\rho \mu} k_{2,\sigma} ) + k_{\mu} (\eta_{\sigma \nu} k_{1,\rho} - \eta_{\rho \nu} k_{1,\sigma} )\right]_3 \ .
\end{multline}
\noindent Here the definition
\begin{equation}
k =  k_1 + k_2
\end{equation}
\noindent was used.

\subsection*{Bubble and triangle vertex contributions with one or more tree level propagators}
This class of diagrams is illustrated in figure \ref{fig:1loopvertswlightconepieces}.  It consists of bubbles and triangles connected to the loop vertices through the order $z^0$ part of the lightcone gauge propagator. Many terms vanish because this part of the propagator is proportional to $q^{\mu} q^{\rho}$, see equation \eqref{eq:AHKgaugeprophardline}. This implies only one or two three vertices with a shifted leg can be sewn onto the loop vertices. 

The contributions of a trivalent triangle vertex with one added lightcone gauge part evaluates to
\begin{multline}\label{eq:loopc}
\epsilon^{\mu}_1 \epsilon_2^{\nu}  A^{(c)}_{\mu\nu\rho\sigma} =  \epsilon^{\mu}_1 \epsilon_2^{\nu}  \int \frac{d^D l}{(2 \pi)^D}  \frac{1}{l^2 (l+k_1)^2(l-k_4)^2}  \left[(D-2) l_{\mu} l_{\sigma} \eta_{\nu\rho} \frac{q^{\alpha} (2 l - k_4)_{\alpha}}{q \cdot k_4} \right]_0  \\
+ 4 \left[ \eta_{\mu\sigma}  \eta_{\nu\rho} \frac{q \cdot l }{q \cdot k_4}   \right]_2
\left(1 \leftrightarrow 2,  3 \leftrightarrow 4, \mu \leftrightarrow \nu, \rho \leftrightarrow \sigma\right) + \mathcal{O} \left(\frac{1}{z} \right)  \ .
\end{multline}

\noindent The bubble vertex with one `lightcone leg' contributes 
\begin{multline}\label{eq:loopd}
\epsilon^{\mu}_1 \epsilon_2^{\nu}  A^{(d)}_{\mu\nu\rho\sigma} =  \epsilon^{\mu}_1 \epsilon_2^{\nu}  \int \frac{d^D l}{(2 \pi)^D}  \frac{1}{l^2(l-k_1 - k_4)^2} \left[ \frac{1}{8} (D-2) \eta_{\mu\sigma} \eta_{\nu\rho} \frac{q^{\alpha} (2 l - k_4)_{\alpha}}{q \cdot k_4} \right]_0 \\ + \left(1 \leftrightarrow 2,  3 \leftrightarrow 4, \mu \leftrightarrow \nu, \rho \leftrightarrow \sigma\right)  + \mathcal{O} \left(\frac{1}{z} \right)  \ ,
\end{multline}
with a $\frac{1}{2}$ symmetry factor included here and in every following bubble. The bubble vertex with two `lightcone legs' contributes
\begin{multline}\label{eq:loope}
\epsilon^{\mu}_1 \epsilon_2^{\nu}  A^{(e)}_{\mu\nu\rho\sigma} = \epsilon^{\mu}_1 \epsilon_2^{\nu}  \int \frac{d^D l}{(2 \pi)^D}  \frac{1}{l^2(l+k_1 + k_4)^2} \left( \left[- \frac{1}{8} (D-2) \eta_{\mu\sigma} \eta_{\nu\rho}  \frac{q^{\alpha} (2 l + k_4)_{\alpha} q^{\beta} (2 l + k_4)_{\beta}}{(q \cdot k_4)^2} \right]_0 \right. \\ +  \left[  \eta_{\mu\sigma} \eta_{\nu\rho} \right]_2+ \mathcal{O} \left(\frac{1}{z} \right) \ .
\end{multline}

\subsection*{Four field bubble, triangle and box one-loop vertices}
This class of diagrams is illustrated in figure \ref{fig:1loopverts4pts}. The four field bubble evaluates to
\begin{multline}\label{eq:loopf}
\epsilon^{\mu}_1 \epsilon_2^{\nu}  A^{(f)}_{\mu\nu\rho\sigma}  = \epsilon^{\mu}_1 \epsilon_2^{\nu}  \int \frac{d^D l}{(2 \pi)^D} \frac{1}{l^2(l+k_1+k_4)^2}  \left(\left[ \frac{1}{8}(D-2)\eta_{\mu\sigma} \eta_{\nu\rho} \right]_0 \right. \\ \left. + \left[(\eta_{\mu\nu} \eta_{\rho\sigma} - \eta_{\mu\rho} \eta_{\nu\sigma})\right]_2 \right) + \mathcal{O} \left(\frac{1}{z} \right) \ .
\end{multline}

\noindent The two (mirrored) four field triangles evaluate to
\begin{multline}\label{eq:loopg}
\epsilon^{\mu}_1 \epsilon_2^{\nu}  A^{(g)}_{\mu\nu\rho\sigma} =  \epsilon^{\mu}_1 \epsilon_2^{\nu}  \int \frac{d^D l}{(2 \pi)^D} \frac{1}{l^2 (l+k_1)^2(l-k_4)^2}  \\ - \left( \left[ (D-2)  l_{\mu} l_{\sigma} \eta_{\nu\rho} \right]_0  
  +\left[2 (\eta_{\mu\rho} \eta_{\nu\sigma} - \eta_{\mu\sigma} \eta_{\nu\rho}  - \eta_{\mu\nu} \eta_{\rho\sigma}) (k_1 \cdot k_4) \right]_{2/3} \right) \\ + 
\left(1 \leftrightarrow 2,  3 \leftrightarrow 4, \mu \leftrightarrow \nu, \rho \leftrightarrow \sigma\right) + \mathcal{O} \left(\frac{1}{z} \right)  \ .
\end{multline}

\noindent  Finally, the four field box evaluates to 
\begin{multline}\label{eq:looph}
\epsilon^{\mu}_1 \epsilon_2^{\nu}  A^{(h)}_{\mu\nu\rho\sigma} = 4 \epsilon^{\mu}_1 \epsilon_2^{\nu}  \int \frac{d^D l}{(2 \pi)^D} \frac{1}{l^2 (l+k_1)^2 (l+k_1+k_2)^2(l-k_4)^2} \left(  \left[\eta_{\rho \sigma} \eta_{\mu\nu} (k_2 \cdot k_3)(k_1 \cdot k_4) \right]_4 \right. \\ \left. + 4 \left[(l_{\mu} l_{\rho} \eta_{\nu\sigma} - l_{\nu} l_{\rho} \eta_{\mu\sigma} )(k_1 \cdot k_4)\right]_2 + (D-2) \left[l_{\mu} l_{\nu} l_{\rho} l_{\sigma}\right]_0 \right) + \mathcal{O} \left(\frac{1}{z} \right) 
\end{multline}
at leading order in $z$. To simplify the second term it was noted that under the loop integral
\begin{equation}
l_{\mu} l_{\nu} \eta_{\rho\sigma} = l_{\rho} l_{\sigma} \eta_{\mu\nu} + \mathcal{O} \left(\frac{1}{z} \right) 
\end{equation}
holds.  Furthermore,
\begin{equation}
(k_1 \cdot k_4) = - (k_2 \cdot k_4) + \mathcal{O} \left(z^0 \right) 
\end{equation}
was used.

\bibliographystyle{JHEP}

\bibliography{loopbib}
\end{document}